\documentclass{article}
\usepackage{graphicx,amsmath,amssymb,amsthm,color,url} 
\usepackage{lineno}

\def\new{\textcolor{black}}

\def\bX{{\bf X}}
\def\bY{{\bf Y}}
\def\bZ{{\bf Z}}
\def\bx{{\bf x}}
\def\by{{\bf y}}
\def\bp{{\bf p}}
\def\R{\mathbb{R}}
\def\V{\mathcal{V}}

\def\E{\mathcal{E}}
\def\Vor{\mathcal{V}or}
\def\Lag{\mathrm{Lag}}
\def\Del{\mathcal{D}el}
\newcommand{\hedge}[2]{\left(#1 \rightarrow #2\right)}
\newcommand{\edge}[2]{\left(#1 - #2\right)}
\newcommand{\eps}{\varepsilon}

\newtheorem{theorem}{Theorem}
\newtheorem{algorithm}{Algorithm}

\theoremstyle{definition}
\newtheorem{definition}{Definition}
\newtheorem{observation}{Observation}

\begin{document}

\title{
  Large-scale semi-discrete optimal transport \\
  with distributed Voronoi diagrams}
\author{Bruno L\'evy$^1$, Nicolas Ray$^2$, Quentin Mérigot$^1$, Hugo Leclerc$^1$ \\
{\small 1. Centre Inria Saclay, Université Paris Saclay, CNRS, Labo. de Maths. d'Orsay} \\
{\small 2. Centre Inria de l'Université de Lorraine, CNRS, LORIA} \\
\\
{\small \tt bruno.levy@inria.fr} \\
{\small \tt nicolas.ray@inria.fr} \\
{\small \tt quentin.merigot@universite-paris-saclay.fr} \\ {\small \tt hugo.leclerc@universite-paris-saclay.fr}}
\date{January, 2025}

\maketitle

\begin{abstract}
In this article, we propose a numerical method to solve semi-discrete optimal transport problems for gigantic pointsets ($10^8$ points and more). By pushing the limits by several orders of magnitude, it opens the path to new applications in cosmology, fluid simulation and data science to name but a few. The method is based on a new algorithm that computes (generalized) Voronoi diagrams in parallel and in a distributed way. First we make the simple observation that the cells defined by a subgraph of the Delaunay graph contain the Voronoi cells, and that one can deduce the missing edges from the intersections between those cells. Based on this observation, we introduce the Distributed Voronoi Diagram algorithm (DVD) that can be used on a cluster and that exchanges vertices between the nodes as need be. We also report early experimental results, demonstrating that the DVD algorithm has the potential to solve some giga-scale semi-discrete optimal transport problems encountered in computational cosmology.
\end{abstract}

\section{Introduction and previous work}

\subsection{Optimal transport in computational physics}

Optimal transport is the study of displacements that minimize a total integrated transport cost --- such as the squared Euclidean distance multiplies by the transported mass, while satisfying a mass conservation constraint. This problem was initially studied by Gaspard Monge in 1784 \cite{Monge1784}, but the first important step towards the mathematical resolution of the transport problem is due to Leonid Kantorovich in the 1940s. A quantum leap in understanding the structure of the problem was made in the 90s by Yann Brenier, with his celebrated polar factorization theorem \cite{BrenierPFMR91}, that revealed rich connections with other fields of mathematics and physics. We refer the reader to \cite{opac-b1122739,OTON,SantambrogioOTsurvey} and references herein.
Another important outcome of optimal transport theory is the definition of transport distances, often referred to as "Wasserstein distances." These metrics enable comparisons between objects of different natures and dimensions, such as point clouds, surfaces, and probability densities. This versatility has been used in numerous recent developments in data science, artificial intelligence and inverse problems. Motivated by the wide spectrum of possible application, significant efforts have been dedicated over the past 20 years to the efficient numerical resolution of optimal transport, resulting in substantial progress
\cite{DBLP:journals/ftml/PeyreC19,DBLP:journals/corr/abs-2003-00855,DBLP:journals/cg/LevyS18}.

In physics, optimal transport has interesting connections with the least action principle \cite{DBLP:journals/nm/BenamouB00} and gradient flows \cite{doi:10.1137/S0036141096303359}. These connections were explored to design new modelling, simulation and inverse methods in various fields such as fluid simulation
\cite{DBLP:journals/focm/GallouetM18,DBLP:journals/tog/GoesWHPD15,DBLP:journals/jcphy/Levy22,DBLP:journals/tog/QuLGJ22}, cosmological reconstruction \cite{EUR,EURNature,10.1093/mnras/stab1676,vhauss_prl_2022,nikak_prl_2022,PhysRevD.108.083534}, alternative theories of gravity \cite{brenier:hal-01137528,levy2024monge}, crowd dynamics \cite{DBLP:journals/dga/BrascoC13,DBLP:journals/nm/BenamouCMO16,DBLP:journals/siamnum/LeclercMSS20} and climatology \cite{MC_GEO_1984,EGAN2022111542,BENAMOU2024112745} to name but a few. \new{However, most numerical implementation of these methods depend on an efficient way of solving the equation that underlies the optimal transport problem, namely the Monge-Ampère equation in the quadratic case. This equation can be viewed as a generalization of the Poisson equation, but solving large-scale instances poses significantly greater challenges. This difficulty is explained by the non-linearity of the Monge-Ampère equation, but also by its degenerate ellipticity, which translates into a convexity constraint that must be satisfied by the solution \cite{benamou2014numerical}.} The Fast Fourier transform had a tremendous impact by making it computationally easy to solve the Poisson equation. Is it possible to find an efficient and scalable computational method that would play the same role for the Monge-Ampère equation and optimal transport ?

A possible candidate is the ``entropic regularization" idea \cite{NIPS2013_af21d0c9}, that uses a softened iterative projection method in dual space, leading to spectacularly efficient solvers \new{for a fixed value of the regularization parameter, which opened the way to many applications of optimal transport in machine learning and data science \cite{DBLP:journals/ftml/PeyreC19}. To enable scaling up to problems of very large size, recent works also explore ways of distributing computations, such as multiple domain decomposition \cite{medina2024flow,mourya2017distributed}, and combination of clever multiresolution techniques (on the point sets) and scaling techniques (on the regularization parameter) to decrease the number of iterations \cite{feydy2019interpolating}. However, despite these improvements, the complexity of each iteration of the Sinkhorn algorithm that is used for solving entropy-regularized optimal transport between unstructured point cloud is quadratic in the size of the problem, prohibiting very large-scale applications.}

In the application domain targeted by the present work, namely computational cosmology, one needs to solve optimal transport problems that are very singular and where the source and target measures are of different nature. The source measure represents the initial condition of the Universe, considered to be uniform\footnote{it is supposed that the Universe had a uniform density at its early ages, which seems to be confirmed by the cosmic microwave background that is very homogeneous} and modeled as the Lebesgue measure over some domain. The target measure, on the other hand, is a finite set, and may contain up to $10^8$ or $10^{10}$ weighted Dirac masses. Each weighted Dirac mass corresponds to a galaxy cluster, modeled as a punctual mass, which makes sense at the cosmological scale.
In these datasets, there are huge variations of density, typically 5 orders of magnitudes or more: there are zones where matter clustered a lot, creating highly dense clusters of galaxies, and other areas that are nearly completely empty of matter. This is illustrated in Figure \ref{fig:laguerre_100M}, p. \pageref{fig:laguerre_100M}). This makes entropy-regularized methods difficult to use, because their efficient implementation depend on regular grids. The cell size is conditioned by the shortest distance between two Dirac masses, which would require in our case extremely fine grid resolution, making space requirement and computation time prohibitive. \new{It may be possible to develop a version working with adaptively subdivided grids, but it is out of reach of current solvers.}

To keep the accuracy required to represent these highly heterogeneous distributions of matter, it is possible to use so-called \emph{semi-discrete} optimal transport \cite{DBLP:journals/cgf/Merigot11,KMT2019,journals/M2AN/LevyNAL15}. Semi-discrete optimal transport exploits the generality
of optimal transport theory, that is written in a general mathematical language (measure theory), that deals not only functions (probability densities), but also much more irregular objects, such as weighted point sets, modeled as measures supported on finite sets. Remarkably, the very same mathematical language can be used from the mathematical modeling to computer implementation, without having mathematical details ``lost in translation", as summarized in the next subsection.

\subsection{From Monge's problem to semi-discrete OT}

\paragraph*{Monge's problem}
We consider our domain $X$ to be a compact subset of the Euclidean space $\R^d$, even though the results that we present below hold on more general spaces, such as tori, which allow to account for periodic boundary conditions. On this set $X$, we consider two non-negative measures $\mu$ and $\nu$ with the same total mass, i.e. $\int_X d\mu$ = $\int_X d\nu$. These measure describe a distribution of some mass in $X$, which might be diffuse when $\mu$ has a density or highly concentrated when, for instance, $\mu$ consists in one or several Dirac masses. We call \emph{transport map between $\mu$ and $\nu$} any map $T:X\to X$ which deforms the mass distribution distributed by $\mu$ onto $\nu$. More formally, this means that for any measureable subset $B$ of $X$, we want amount of mass of $\nu$ contained in $B$ to be equal to the amount of mass of $\mu$ contained in the pre-image $T^{-1}(B)$, i.e. $\mu(T^{-1}(B)) = \nu(B)$. If $T$ is a transport map between $\mu$ and $\nu$, we denote $T_\#\mu = \nu$. The quadratic version of Monge's problem
\cite{Monge1784} consists in finding a transport map $T: X \rightarrow X$
that minimizes the transport cost:
\begin{equation}\label{eqn:Monge}
\inf_{T:X\to X\mid T_{\#}\mu = \nu} \int_{X} \| T(\bx) - \bx \|^2 d\mu(\bx)
\end{equation}
We note at this point that this problem can be very ill-posed, and in particular that the set of admissible transport maps might be empty.

\color{black}
\paragraph*{Kantorovich's relaxation}

Most numerical methods for solving optimal transport rely on (the dual of) a relaxed version of Monge's problem, which was proposed by Kantorovich in the 1940s. The relaxation is given by
\begin{equation} \label{eqn:Kanto1}
\inf_{\gamma \in \Gamma(\mu,\nu)}  \int_{X \times X} \| \bx - \by \|^2 d\gamma(\bx,\by)
\end{equation}
where the set $\Gamma(\mu,\nu)$ is the set of non-negative measures over $X\times X$ with marginals $\mu$ and $\nu$, that is, satisfying the constraints
$$  \Pi_1 \# \gamma = \mu,\qquad  \Pi_2 \# \gamma= \nu$$
where $\Pi_i: (x_1,x_2) \in X\times X \to x_i \in X$ are the projection maps.
A measure $\gamma$ that
satisfies these two marginal constraints is called a \emph{transport plan} between $\mu$ and $\nu$. It corresponds to the notion of coupling between the measures $\mu$ and $\nu$ in probability theory.

The measure $\gamma$ is supported on the Cartesian product $X \times X$ and may be thought of as a possibly diffuse approximation of the graph of a transport map. The quantity $\gamma(A\times B)$ is the amount of mass that is transported from the set $A\subset X$ to the set $B\subset X$ by the transport plan $\gamma$. The marginal contraint $\Pi_{1}\#\gamma = \mu$ implies that $\gamma(A\times X) = \mu(A)$ and encodes the fact that all the mass emanating from the set $A$ has to be sent by $\gamma$ to some point in $X$. The constraint $\Pi_2\#\gamma = \nu$ is interpreted similarly.

In the relaxed problem \eqref{eqn:Kanto1} the objective function and the constraints are both linear in the unknown transport plan $\gamma$, i.e. Kantorovich's relaxation is an infinite-dimensional linear programming problem.

\color{black}
\paragraph*{Kantorovich dual and c-transform}

The linear and linearly constrained nature of Kantorovich's problem calls for taking a look at the dual formulation:
\begin{equation}\label{eqn:KantoDual}
\begin{array}{l}
\sup_{\phi,\psi} \left[ \int_X \phi(\bx) d\mu + \int_X \psi(\by) d\nu \right] \\[3mm]
  \mbox{subject to:} \quad \phi(\bx) + \psi(\by) \le \| \bx - \by \|^2 \quad \forall \bx,\by \in X \times X
\end{array}
\end{equation}
where $\phi$ and $\psi$, the Lagrange mutlipliers associated with the marginal constraints, are both functions from $X$ to $\R$. It is interesting to notice that as compared to the primal problem, that depends on an unknown transport plan $\gamma$ on $X \times X$, the dual problem depends on two unknown functions $\phi, \psi: X \rightarrow \R$. Assuming that the measures $\mu$ and $\nu$ are probability densities, it is known since the work of Brenier \cite{BrenierPFMR91} that Monge's problem \eqref{eqn:Monge} admits a solution $T$. The optimal transport map $T$ from $\mu$ to $\nu$ and its inverse $T^{-1}$ are connected to the maximizers of the dual Kantorovich problem through the following formulas:
\begin{equation}
 \begin{array}{lcl}
    T(\bx)      & = & \bx - \frac12\nabla \phi(\bx) \\[2mm]
    T^{-1}(\by) & = & \by - \frac12\nabla \psi(\by)
 \end{array}
 \label{eqn:OTM}
\end{equation}
With computer implementation in mind, considering the dual problem seems to offer significant advantages: it appears much easier to work with two functions, $\phi, \psi: X \to \mathbb{R}$, rather than a measure $\gamma$ supported on $X \times X$. If the space $X$ is discretized using $N$ points, the dual problem approach involves \(2N\) variables, each function being represented as a vector with $N$ entries, compared to the $N^2$ variables potentially required to discretize the transport plan $\gamma$. \new{However, in this formulation of the dual problem, the number of constraints becomes quadratic. This challenge can be addressed using the concept of the $c$-transform, a variant of the concept of convex conjugation or Legendre-Fenchel transform. Specifically,\footnote{The problem is symmetric, one could swap the roles of $\phi$ and $\psi$. Here we consider that the second Lagrange multiplier $\psi$ is given and that the first one $\phi$ is deduced as the $c$-transform of $\phi$, in that order. As we shall see further on, in the semi-discrete setting, $\psi$ has a finite support and can be represented as a vector of $N$ values.} if $\psi: X \to \mathbb{R}$ is given, the function $\phi$ that maximizes the dual cost $\int \psi \, d\mu$ while satisfying the constraints $\phi(\bx)+\psi(\by)\leq \Vert{\bx-\by}\Vert^2$ can be constructed easily: one should take $\phi = \psi^c$, where}
\begin{equation}\label{eqn:ctransform}
\psi^c(\bx) = \inf\limits_{\by \in X} \| \bx - \by \|^2 - \psi(\by).
\end{equation}
Using this construction of the $c$-transform, the dual problem becomes the following problem, that depends on a single function $\psi: X \rightarrow \R$:
\begin{equation}\label{eq:KantoThird}
\sup_{\psi} \mathcal{K}(\psi) = \int_X \psi^c(\bx) d\mu + \int_X \psi(\by) d\nu
\end{equation}
\new{From an implementation viewpoint, the main advantage of this formulation is that we have turned the (constrained) linear programming problem \eqref{eqn:KantoDual} into an \emph{unconstrained} and concave maximization problem. We have however, moved the difficulty induced by constraints in \eqref{eqn:KantoDual} into the computation of the $c$-transform $\psi^c$ in the unconstrained problem \eqref{eq:KantoThird}.}

\color{black}
\paragraph*{Semi-discrete optimal tranport} Up to now we did not make any assumption
regarding the measures $\mu$ and $\nu$ besides the assumptions that they have the same total mass.
Semi-discrete optimal transport is a special case, where
$\mu$ is a density over the set $X$ and $\nu = \sum_{i=1}^N \nu_i \delta_{\bx_i}$ is a finitely supported measure, i.e. a weighted sum of Dirac masses, satisfying the mass balance $\sum_i \nu_i = \int_X d\mu$. We also assume that the points $\bx_i$ are distinct. In this particular setting, the Kantorovich dual \eqref{eq:KantoThird} can be further simplified. Indeed, a function $\psi$ on the finite support of the measure $\mu$ can be fully described by its values at the points $\bx_i$, denoted $\psi_i$, thus by a vector of $N$ scalar values $(\psi_i)_{1\leq i\leq N}$. We therefore consider the function $K:\mathbb{R}^N\to\mathbb{R}$ defined by
\begin{equation*}
\begin{aligned}
    K(\psi) & = \int_X \psi^c(\bx) d\mu(\bx) + \sum_{i=1}^N \psi_i \nu_i \\[2mm]
            & = \int_X \min\limits_{i} \left[ \| \bx - \bx_i \|^2 - \psi_i  \right] d\mu(\bx) + \sum_i \psi_i \nu_i \\[2mm]
\end{aligned}
\end{equation*}
Since the minimum inside the integral is taken over a finite set, it is  very natural to introduce the minimization diagram of the functions $\| \bx - \bx_i \|^2 - \psi_i$, which is often called a Laguerre diagram in computational geometry. More precisely, we define the $i$th Laguerre cell to be
$$ \Lag_i^\psi = \left\{
\bx \in \mathbb{R}^d \mid \| \bx - \bx_i \|^2 - \psi_i \ \le\ \| \bx - \bx_j \|^2 - \psi_j
\quad \forall j \neq i \right\}.$$
With this definition at hand, one might replace the integral over $X$ in \eqref{eqn:SD1} into a sum of integral over Laguerre cells:
\begin{equation}\label{eqn:SD1}
  K(\psi) = \sum_i \int_{\mbox{Lag}^\psi_i} \left( \| \bx - \bx_i \|^2 - \psi_i \right) d\bx + \sum_i \psi_i \nu_i
\end{equation}
In the setting that we just described, the Kantorovich dual $K$ depends on the vector $(\psi_i)_{1\leq i\leq N}$. The Kantorovich potential $\phi$, on the other hand, is continuous and defined on the whole domain $X$. However, it is defined as the $c$-transform $\psi^c$ of $\psi$, and is therefore also naturally parameterized by the same vector. We summarize some of the known results about this functional and its relation to semi-discrete optimal transport into a theorem, which follows from \cite[\S5]{aurenhammer1998minkowski}, and whose proof can be found in \cite[Theorem 40]{DBLP:journals/corr/abs-2003-00855}.

\begin{theorem}
    Assume that all the points $\bx_1,\hdots,\bx_N$ are in distinct. The function $K: \mathbb{R}^N\to\mathbb{R}$ is concave, continuously differentiable, and its gradient is given by
    \begin{equation} \label{eqn:g}
    \frac{\partial K}{\partial\psi_i} = \nu_i - \int_{\Lag_i^\psi} d\mu.
    \end{equation}
    Moreover, $K$ admits at least a global maximizer. If $\psi$ is this maximizer and $\phi(\bx) = \psi^c(\bx) = \min_i \Vert\bx - \bx_i\Vert^2 - \psi_i$ its $c$-transform, the optimal transport map between $\mu$ and $\nu$ is given by $T: \bx\mapsto\bx - \frac{1}{2}\nabla \phi(\bx)$. This map $T$ is piecewise constant, equal to $\bx_i$ on the $i$th Laguerre cell $\Lag_i^\psi.$
\end{theorem}

\newcommand{\Adm}{\mathrm{Adm}}
\newcommand{\vol}{\mathrm{vol}}
Thanks to this theorem, we could think of solving the dual of the Kantorovich problem using gradient ascent of the functional $K$, as suggested in \cite[\S5]{aurenhammer1998minkowski}. However, it turns out that because of the special structure of the functional $K$, evaluating $K(\psi)$ or $\nabla K(\psi)$ is as costly as evaluating the second derivative $D^2 K(\psi)$. We will therefore rely on a second order Newton method for the maximization of $K$ rather than a first-order method. To do so, we will restrict the set of potentials $\psi$ that are admissible in the iterations of the algorithm. We will denote
$$\Adm_\eps = \left\{ \psi\in\mathbb{R}^N \mid \forall i, \int_{\Lag_i^\psi}d\mu \geq \eps \right\}, \quad \Adm = \bigcup_{\eps>0} \Adm_\eps. $$
Thus, $\Adm$ is the set of potentials $\psi$ such that all Laguerre cells contains a positive fraction of the mass of $\mu$. Under some regularity assumptions on $\mu$ and $X$ one can show that $K$ is twice differentiable and give an explicit expression for its second derivatives. The following theorem is a special case of
\cite{KMT2019}, see also \cite{journals/M2AN/LevyNAL15}. It shows that the functional $K$ is twice continuously differentiable on the admissibility set $\Adm$:

\begin{theorem}
Assume that $\mu$ has a continuous density on a convex compact domain $X$, which is bounded from below by a positive constant. Then, $K$ is twice continuously differentiable on $\Adm$ and its second partial derivatives are
\begin{equation} \label{eqn:H}
\begin{aligned}
    \frac{\partial^2 K}{\partial \psi_i \partial \psi_j}(\psi) & =
    \frac{1}{2} \frac{1}{\| \bx_j -  \bx_j\|}
    \int_{\Lag_{i,j}^\psi} \mu(x) d \vol^{d-1}(x)  \quad \mbox{if } j \neq i \\
    \frac{\partial^2 K}{\partial \psi_i^2}(\psi) & =  - \sum_{j \neq i} \frac{\partial^2 K}{\partial \psi_i \partial \psi_j}(\psi)
\end{aligned}
\end{equation}
where $\Lag^\psi_{i,j}$ denotes the $(d-1)$-dimensional intersection between the $i$th and $j$th Laguerre cells, and where the integral in the first formula should be interpreted as an integral with respect to the surface measure.
\end{theorem}

\color{black}
We note that the formula \eqref{eqn:H} is reminiscent of a finite-volume discretization of the Poisson equation on the mesh defined by the Laguerre cells \cite[\S3]{eymard2000finite}, which may not be surprising as the equation $\nabla K(\psi) = 0$ is a discretized Monge-Ampère equation, as explained in \cite[Remark 18]{DBLP:journals/corr/abs-2003-00855}. From formula \eqref{eqn:H}, we can directly see that the matrix $D^2 K(\psi)$ of second-order partial derivatives is weakly diagonally dominant with negative diagonal entries, and therefore has non-positive eigenvalues. We could also have derived this property from the concavity of $K$. Exploiting in addition the connectedness of the graph over the set $\{1,\hdots,N\}$, where two vertices $i\neq j$ are connected if $D^2 K(\psi)_{i,j} > 0$, we can verify that the kernel of the matrix $D^2 K(\psi)$ is spanned by the constant vector $(1,\hdots,1)$. This shows that the restriction of $K$ on $\Adm \cap \{\psi \in {\mathbb R}^N \ |\ \psi_0 = 0\} $ is \emph{strongly concave}\footnote{where the constraint $\psi_0 = 0$
(for instance) determines the translational degree of freedom along the $(1,\ldots,1)$ vector. Intuitively, the vector of Lagrange
multipliers $\psi$ may be thought of as a \emph{potential}, defined up to a translational constant $(C,\ldots,C)$.
}.
\color{black}



At this point, we have done nothing else than rewriting the generic optimal transport problem in the specific case where $\mu$ is the Lebesgue measure and $\nu$ is an empirical measure. Let us now explain how to compute the optimal transport in practice, that is, how to determine the vector $(\psi_i)_{1\leq i\leq N}$.
The Kantorovich dual that depends on these parameters is concave, twice differentiable and strongly convex on $\Adm$. Moreover, the solution $\psi$ is characterized by $\int_{\Lag_i^\psi} d \mu = \nu_i > 0$, and therefore also belongs to this set $\Adm$. The Newton method that we will use will rely on a specific backtracking which ensures that its iterates remain in the set $\Adm$, as explained in Algorithm~\ref{algo:KMT}. More specifically, the algorithm iteratively halves the descent parameter $\alpha$ until two criteria are met: the volume of the smallest cell needs to be larger than a threshold $a_0$ (first condition in line 3), and the norm of the gradient needs to decrease sufficiently (second condition in line 3). The threshold $a_0$ for the minimum cell volume corresponds to (half) the minimum cell volume for $\psi = 0$ (also called Voronoi diagram) and minimum prescribed area $\nu_i$.

\begin{algorithm} Semi-discrete optimal transport
$$
\begin{array}{l}
    \begin{array}{ll}
    \mbox{\bf input:}  & \mbox{\rm a pointset } (\bx_i)_{i=1}^N\ \mbox{\rm and masses } (\nu_i)_{i=1}^N\\
    \mbox{\bf output:} & \mbox{\rm the Laguerre diagram } \{\mbox{\rm Lag}_i^\psi\}_{i=1}^N \mbox{\rm such that } | \mbox{\rm Lag}_i^\psi | = \nu_i \ \forall i
    \end{array} \\[4mm]
    \hline\\
    \begin{array}{ll}
 (0) & \psi \leftarrow 0 \\
 (1) & \mbox{\bf while} \| \nabla K \|_\infty < \epsilon \\
 (2) & \quad \mbox{\bf solve for } \bp \ \mbox{\rm in }
    [\nabla^2 K(\psi)] \bp = - \nabla K(\psi) \\
 (3) & \quad \mbox{\rm find descent parameter } \alpha\ \mbox(\rm see\ Algorithm\ \ref{algo:KMT})\\
 (4) & \quad \psi \leftarrow \psi + \alpha \bp \\
 (5) & \mbox{\bf end while}
    \end{array}
\end{array}
$$
\label{algo:SDiscrete}
\end{algorithm}

\begin{algorithm}  Kitagawa-Mérigot-Thibert descent (KMT)
$$
\begin{array}{l}
    \begin{array}{ll}
    \mbox{\bf input:}  & \mbox{\rm current values of } (\psi_i)_{i=1}^N\ \mbox{\rm and Newton direction } \bp\\
    \mbox{\bf output:} & \mbox{\rm descent parameter } \alpha \
    \mbox{\rm determining the next iterate }  \psi \leftarrow \psi + \alpha \bp
    \end{array} \\[4mm]
    \hline\\
    \begin{array}{ll}
    (1) & \alpha \leftarrow 1 \\
    (2) & \mbox{\bf loop} \\
    (3) & \quad \mbox{\bf if  } \inf_i | \mbox{\rm Lag}^{\psi + \alpha \bp}_i | >a_0
     \quad \mbox{\bf and} \quad \| \nabla K(\psi + \alpha \bp) \| \le (1 - \alpha / 2) \| \nabla K(\psi) \| \\
    (4) & \quad \quad \mbox{\bf then exit loop} \\
    (5) & \quad \alpha \leftarrow \alpha / 2 \\
    (6) & \quad \mbox{\rm Compute the Laguerre diagram } (\mbox{\rm Lag}^{\psi+\alpha \bp}_i)_{i=1}^N \\
    (7) & \mbox{\bf end loop} \\
     & \mbox{\rm where } a_0 = \frac{1}{2}\min\left( \inf_i  \left|\mbox{\rm Lag}^{\psi=0}_i\right|  , \inf_i(\nu_i) \right).
    \end{array}
\end{array}
$$
\label{algo:KMT}
\end{algorithm}

In the numerical algorithm for semi-discrete optimal transport above, the two dominant costly operations are computing Laguerre diagrams and solving linear systems. By examining the coefficients of the linear system given in Equation \ref{eqn:H}, we have already noticed a strong similiarity with a Poisson system. {\textcolor{black}{Like in a Poisson system, the Hessian $\nabla^2 K$ is semi-definite positive. It has the vector $[1,1 \dots 1]^t$ in its kernel. In such a configuration, the conjugate gradient algorithm converges to the unique solution of the system orthogonal to the kernel, which is also the minimum-norm one (see e.g. \cite{hayami2020convergenceconjugategradientmethod})}}. To accelerate the convergence of the conjugate gradient algorithm, an algebraic multigrid pre-conditioner \cite{Demidov2019} is highly effective, as we noticed in \cite{levy2024monge} and confirmed experimentally. In this setting, the lion's share of the computational cost is the computation of Laguerre diagrams: for a typical reconstruction problem in cosmology, it takes 60\% of the computation time (whereas matrix assembly and linear solve take 25\% and 15\% respectively). Besides computational cost, another important obstacle to scaling up is the memory cost of the data structure used to store the Laguerre diagrams. Both aspects call for a distributed algorithm, that could run on a cluster. We shall now review the standard algorithm to compute (generalized) Voronoi diagrams, then we will see how to derive a distributed version, that can run on a cluster.


\subsection{Voronoi diagrams and algorithms to compute them}

Voronoi diagrams are ubiquitously used in various domains of computational sciences, including geometry processing, data analysis, machine learning and computational physics to name but a few.
The properties of Voronoi diagrams and their various generalizations were studied
\cite{10.1145/116873.116880,10.1145/77635.77639} and used to derive efficient computational
algorithm, see surveys in \cite{DBLP:books/daglib/0095173} and \\
\noindent\cite{Okabe00}.
To efficiently represent a Voronoi diagram in a computer, one can use the dual combinatorial structure,
called the Delaunay triangulation.
Historically, one of the first methods, proposed in the 80s by Sloan \cite{SLOAN198734}, transforms an
arbitrary triangulation into the Delaunay triangulation by flipping the edges until all the triangles
are valid, that is, such that their circumcircles are empty of any other point.
The most widely used algorithm to compute Voronoi diagram was proposed in the 80s, simultaneously by Bowyer and
Watson \cite{DBLP:journals/cj/Bowyer81,journals/cj/Watson81}. It also operates on the Delaunay triangulation.
It inserts the points one by one in the triangulation,
and iteratively discards the triangles that violate the empty circumcircle condition and creates new ones, connecting
the newly inserted points to the boundary of the zone where triangles were discarded, which is more efficient than
flipping edges. \\

Numerical precision is an important aspect to be considered with uttermost care when implementing these algorithms.
The most important and subtle part of a geometric algorithm is a set of functions, called geometric predicates, that take as argument geometric objects (for instance two points $\bx_1$ and $\bx_2$) and that
returns a discrete value (for instance, +1 if $\bx_1$ is above $\bx_2$, -1 if $\bx_1$ is below $\bx_2$ and 0 if they
have the same altitude). In general, these geometric predicates are more complicated, and correspond to the sign of a polynomial in the points coordinates. The main difficulty is that the floating point numbers manipulated in computers have limited precision. If no care is taken, the computed signs are not exact, and depending on the order of the operations, the predicate may indicate that $\bx_1$ is above $\bx_2$ in one part of the algorithm, and that $\bx_1$ is below $\bx_2$ in another part, leading to inconsistencies in the constructed combinatorial structure. For this reason, different techniques were
proposed to compute the exact sign of the polynom, and several software packages were proposed, by Shewchuk \cite{shewchuk97a}
(who uses arithmetic expansions), in CGAL \cite{cgal:eb-23b} (that uses a combination of arithmetic filters, interval arithmetic and exact numbers),PCK \cite{DBLP:journals/cad/Levy16} and GEOGRAM \cite{WEB:GEOGRAM}, that use a combination of arithmetic filters and arithmetic expansions. \\

  To speed-up the computation of Voronoi diagrams, a parallel algorithm based on the ``security radius" theorem was proposed
  \cite{DBLP:conf/imr/LevyB12}, as well as a version that runs on GPUs for Voronoi and Laguerre diagrams \cite{DBLP:journals/tog/RayS0L18,DBLP:journals/cgf/BasselinARSLL21}.
To compute Delaunay triangulations and Voronoi diagram with gigantic pointsets, several techniques were proposed, based on
decomposing the input domain into regions. The ``streaming" algorithm proposed in \cite{DBLP:journals/tog/IsenburgLSS06} inputs a stream of points and outputs a stream of triangles. It minimizes the globally stored information based on ``finalization" tags inserted in the point stream, indicating that a point is the last one in its region, which allows one to discard the information that is no-longer needed.
Another approach consists in first computing the Delaunay triangulation in each region independently, then determine the points that
should be exchanged with the neighboring regions in order to get the correct triangulation / Voronoi cells \cite{7013068}. We mention also the AREPO code \cite{10.1111/j.1365-2966.2009.15715.xISTEX}, used for large scale cosmological simulations, that is based on a variant of Sloan's edge flipping algorithm, distributed over a cluster, using an algorithm similar to the one in \cite{7013068}. As done in \cite{DBLP:conf/bigdataconf/CaraffaMYB19}, a similar result can be obtained by using the \emph{star splaying} algorithm \cite{10.1145/1064092.1064129}, that can restore the Delaunay condition in a nearly-Delaunay triangulation. It represents a mesh as a collection of (possibily incoherent) vertex stars, and restores coherency between them. \\

As explained later, the method proposed in this article shares some similarities with the \emph{stars splaying} idea. The main difference is the path of reasoning that we followed, that is nearly equivalent, but that is structured around the Voronoi cells (instead of the
1-ring neighborhoods). This path of reasoning naturally leads to several algorithms for parallel and distributed Voronoi diagrams. As fas as the distributed algorithm is concerned, the approach results in a smaller number of points exchanged between the regions. \\

With the goal of making it possible for semi-discrete optimal transport (see Section \ref{sec:OT}) to scale up beyond $10^8$ particles, what follows introduces new algorithms to compute (generalized) Voronoi diagrams in parallel and in a distributed way. The idea is based on the simple observation that the cells defined by a subgraph of the Delaunay graph contain the Voronoi cells, and that their intersections define the edges of the Delaunay graph (Section \ref{sec:basics}). From this observation, we deduce two algorithms, a parallel one (PVD for Parallel Voronoi Diagram), described in Section \ref{sec:PVD}, that can be used in multicore shared memory machines,
and a distributed one (DVD for Distributed Voronoi Diagram), described in Section \ref{sec:DVD}, that computes a (potentially gigantic) Voronoi diagram on a cluster, and exchanges vertices between the nodes as need be.
The distributed Voronoi Diagram algorithm (DVD) can be used to design a giga-scale semi-discrete optimal transport algorithm (see Section \ref{sec:OT} for early results).

\section{Distributed Voronoi Diagram}
\subsection{The Voronoi diagram}
\label{sec:basics}

\begin{figure}
    \centering
    \includegraphics[width=\textwidth]{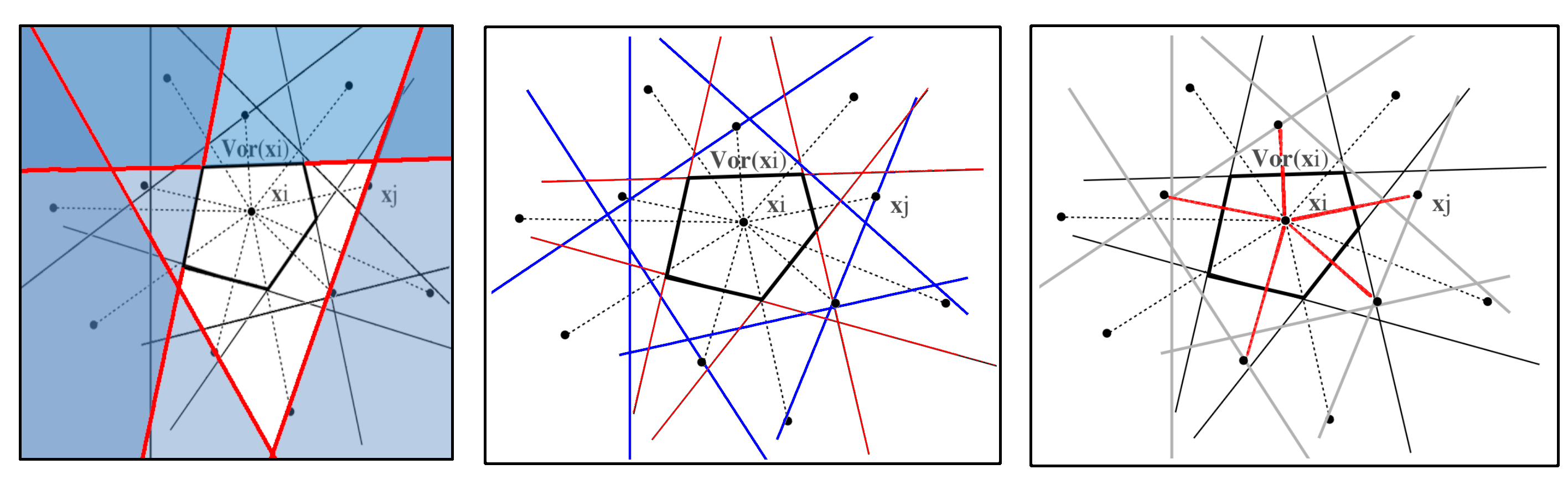}
    \caption{
     Left: the intersection of four half-spaces $\Pi^+(i,j)$ (their boundaries are highlighted in red). The area shaded in blue corresponds to the excluded half-spaces. The white region corresponds to the intersection between four half-spaces. It contains the Voronoi cell.
     Center: The Voronoi cell $\Vor(\bx_i)$ corresponds to the intersection of all the half-spaces
      $\Pi^+(i,j)$. Among them, some touch the boundary of $\Vor(\bx_i)$ and are contributing (in red), and some are non-contributing (in blue). Right: the Delaunay graph (in red) corresponds to the set of edges $\edge{i}{j}$ such that $\Pi^+(i,j)$ is contributing to $\Vor(\bx_i)$ (or equivalently, $\Pi^+(j,i)$ is contributing to $\Vor(\bx_j)$).
    }
    \label{fig:voro_clip}
\end{figure}

\color{black}
\begin{definition}[Voronoi diagram]\label{def:Voronoi}
The \emph{generalized Voronoi diagram} $\Vor(F)$, also called \emph{lower envelope} or \emph{minimization diagram}, associated to a set of $N$ continuous functions $F_1,\hdots,F_N$ on $\R^d$ is the set of
\emph{generalized Voronoi cells} defined by
 $$
    \Vor_i(F) = \{ \bx \in\R^d | \forall j \neq i,~\ F_i(\bx) \le F_j(\bx) \}.
 $$
The $i$th generalized Voronoi cells can be recovered by intersecting the $N-1$ \emph{dominance regions} $\Pi^+_{i,j}(F)$ for $j\neq i$ where
\begin{equation}\label{eqn:vor}
    \Pi^+_{i,j}(F) = \{ \bx \in\R^d \ |\ F_i(\bx) \le F_j(\bx) \}.
\end{equation}
Finally we call \emph{bisector} the set of points $\Pi(i,j)(F) = \left\{ \bx\ |\ F_i(\bx) = F_j(\bx) \right\}$.
\end{definition}

If $\bx_1,\hdots,\bx_N$ are $N$ distinct points in $\R^d$ (called \emph{generating points} or \emph{sites}), and if $F_i(\bx) = \|\bx - \bx_i\|^2$, then the generalized Voronoi cell $\Vor_i(F)$ coincides with the usual Voronoi cell associated with the point $\bx_i$. If in addition we are given $N$ scalars $\psi_1,\hdots,\psi_N$ and  if $F_i(\bx) = \|\bx - \bx_i\|^2 - \psi_i$, then the minimization diagram is called a \emph{Laguerre} or \emph{power} diagram. In both cases, the dominance regions are halfspaces and the general Voronoi cells are convex polyhedra. These notions are illustrated in  Figure \ref{fig:voro_clip}.

\begin{definition}[Voronoi diagram associated to a graph]
We call a \emph{graph} over $\{1,\hdots,N\}$ any subset $\E$ of the product space $\{1,\hdots,N\}^2$. An edge $i\to j$ belongs to the graph if and only if  the pair $(i,j)$ belongs to $\E$. The Voronoi cell associated to the graph $\E$ is defined by:
\begin{equation} \label{eqn:V}
    \Vor^\E_i(F) = \bigcap\limits_{j \hbox{ s.t } (i,j)\in\E} \Pi^+_{i,j}(F),
\end{equation}
\end{definition}
and the Voronoi diagram $\Vor^\E(F)$ associated to the graph $\E$ is the set of cells $\Vor_i^\E(F)$.

One can make the following (trivial yet useful) observation:
\begin{observation} \label{obs:cells}
Given an arbitrary set $\E$ of oriented edges, the Voronoi regions with respect to the graph $\E$ contain the corresponding Voronoi regions:
$$
    \Vor_i(F) \subseteq \Vor^\E_i(F)
$$
In general, the inclusion is strict, and the regions $\Vor_i^\E(F)$ may overlap (not only over their boundaries).
\end{observation}
\begin{proof}
    This property is directly deduced from the definition of $\Vor_i(F)$ and $\Vor_i^\E(F)$:
    \begin{equation*}
        \Vor_i(F) = \bigcap_{j\neq i} \Pi_{i,j}^+(F) \subseteq \bigcap_{j\mid (i,j)\in \E} \Pi_{i,j}^+(F). \qedhere
    \end{equation*}
\end{proof}


In terms of these definitions, starting from an arbitrary oriented graph $\E$, our goal is to find a new graph $\E^\prime$ that contains all the edges that are missing in $\E$ for $\Vor^{\E^\prime}(F)$ to be the Voronoi diagram $\Vor(F)$.
To do so, we make another simple observation:
\begin{observation}\label{obs:subgraph}
Let $\E$ be a graph over $\{1,\hdots,N\}$, and define
$$ \E' = \{ (i,j) \in \{1,\hdots,N\}^2\mid \Vor_i^\E(F) \cap \Vor_j^\E(F) \neq \emptyset\}, $$
Then, $$\Vor^{\E'}(F) = \Vor(F).$$
\end{observation}

\begin{proof}
   We already know that the cell $\Vor_i^{\E'}(F)$ relative to the graph $\E'$ contains the Voronoi cell $\Vor_i(F)$.
   We now need to prove inclusion in reverse order.
   Consider a point $\bx$ in $\Vor_i^{\E'}(F)$. We shall show that $F_i(\bx)$ is smaller than $F_j(\bx)$ for all $j$'s:

\begin{itemize}
   \item  For an edge $(i,j)$ of the graph $\E'$ we have,
   by definition of $\Vor_i^{\E'}(F)$:
   $$ \forall j  \in \{1, \ldots, N\}, (i,j) \in \E' \Rightarrow F_i(\bx) \le F_j(\bx).$$

 \item Consider now an edge $(i,j)$ that is not in $\E'$. By definition of $\E'$, we have
   $\Vor_i^{\E'}(F) \cap \Vor_j^{\E'}(F) = \emptyset$, thus
   $\Vor_i^{\E'}(F) \cap \Vor_j(F) = \emptyset$ (because $\Vor_j(F) \subseteq \Vor_j^{\E'}(F)$), hence there is no Voronoi
   cell $\Vor_j(F)$ that contains $\bx$:

   $$
   \begin{array}{l}
   \forall j\ \in \{1,\ldots,N\} \setminus {\cal N}_i^{\E'},\ \bx \notin \Vor_j(F) \\[2mm]
   \mbox{where:} \quad {\cal N}_i^{\E'} = \{ i \} \cup \{ j\ |\ (i,j) \in \E' \}.
   \end{array}
   $$

   Remembering that
   the Voronoi diagram is a covering of ${\mathbb R}^d$, there exists a Voronoi cell $\Vor_k(F)$ that contains $\bx$, with
   $k$ in the complement of $\{1,\ldots,N\} \setminus {\cal N}_i^{\E'}$, that is,
   $k \in {\cal N}_i^{\E'}$. Hence,
   $F_j(\bx) \ge \inf_{k \in N_i^{\E'}} F_k(\bx)$.
   Remembering that $\bx$ is in $\Vor_i^{\E'}(F)$ and by definition of $\Vor_i^{\E'}(F)$, we have
   $\inf_{k \in N_i^{\E'}} F_k(\bx) = F_i(\bx)$ hence $F_j(\bx) \ge F_i(\bx)$.
\end{itemize}

\end{proof}
\color{black}

\subsection{PVD: Parallel Voronoi Diagram}
\label{sec:PVD}
\begin{figure}
    \centerline{
       \includegraphics[width=\textwidth]{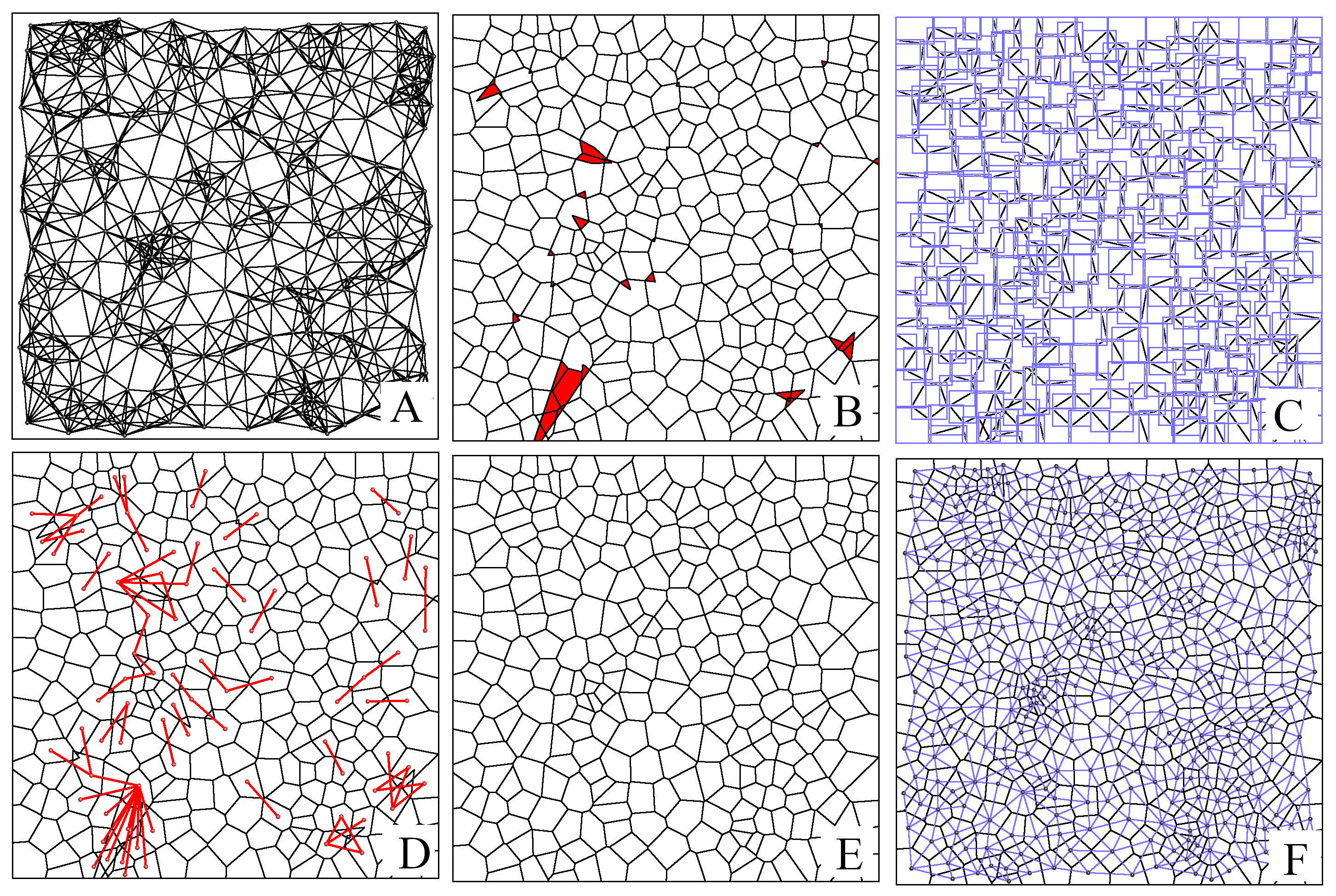}
    }
    \caption{ A: k-NN graph; B: the cells deduced from the k-NN graph. They mostly correspond to the Voronoi diagram, except some intersections (in red); C: detecting the intersections with an Axis-Aligned Bounding Box tree; D: adding an edge in the graph for each pair of intersecting cells; E: the cells now
    correspond to Voronoi cells; F: the Delaunay graph contains the edges that correspond to the contributing bisectors. One gets the Voronoi cells as soon as the graph contains all these edges (as is the case at step D).
    }
    \label{fig:PVD}
\end{figure}

From Observation \ref{obs:subgraph}, one can deduce a new algorithm to compute a Voronoi diagram:
\begin{algorithm} Parallel Voronoi Diagram (PVD)
$$
\begin{array}{l}
    \begin{array}{ll}
    \mbox{\bf input:}  & \mbox{\rm a pointset } \bX \\
    \mbox{\bf output:} & \mbox{\rm the Voronoi diagram } \left\{\Vor_i\right\}_{i=1}^N
    \end{array} \\[4mm]
    \hline\\
    \begin{array}{ll}
    (1) & \E \leftarrow kNN(\bX) \\[2mm]
    (2) & \mbox{\rm compute the cells } (\V^\E_i)_{i=1}^N \\[2mm]
    (3) & \E \leftarrow \E \cup \left\{ \edge{i}{j}\ |\ \V^\E_i \cap \V^\E_j \neq \emptyset \right\} \\[2mm]
    (4) & \mbox{\rm re-compute the cells } (\V^\E_i)_{i=1}^N \\[2mm]
    \end{array}
\end{array}
$$
\label{algo:PVD}
\end{algorithm}
where:
\begin{itemize}
    \item in step (1), the k-nearest neighbor graph ($kNN$) is extracted from a kD-tree (Figure \ref{fig:PVD}-A). Both the kD-tree construction and extraction of the k-nearest neighbors is computed in parallel. \textcolor{black}{The reader is referred to \cite{10.5555/77589} for a survey on spatial data structures (including kNNs, KD-trees, and axis-aligned bounding box trees used later in step 3). The number $k$ of neighbors in the $kNN$ is chosen based on the average number of neighbors in a Delaunay triangulation (6 in 2D, 16 in 3D)};
    \item in step (2), the cells $V^\E_i = \bigcap_{\hedge{i}{j}\in\E} \Pi^+(i,j)$ are computed in parallel. In the case
       of a Voronoi diagram or Laguerre diagram, the dominance regions $\Pi^+(i,j)$ are half-spaces. Their intersection are
       efficiently computed by representing the cells in dual form \cite{10.1145/116873.116880,10.1016/j.jcp.2021.110838}.
       The cells computed at step (2) are shown in Figure \ref{fig:PVD}-B. As one can see, the set of cells $\V^\E_i$
       deduced from the k-NN graph $\E$ is very near the Voronoi diagram, except for a few overlaps between the cells (in red);
    \item in step (3), we use an axis-aligned bounding box tree to determine a set of candidate intersection pairs
       (Figure \ref{fig:PVD}-C). Then all the
       candidate intersections are processed in parallel. Each time the bounding boxes of a pair of cells $\V^\E_i$ and $\V^\E_j$ have an intersection, the edges $\edge{i}{j}$ and $\edge{j}{i}$ are appended to $\E$ (in red in Figure \ref{fig:PVD}-D). This will insert
       too many edges, but any supergraph will give the correct result;
    \item in step (4), each re-computed cell $V_i^\E$ corresponds to the Voronoi cell $\Vor_i$ (since $\Del(\bX) \subseteq \E$). The so-computed Voronoi diagram is shown in Figure \ref{fig:PVD}-E. The Delaunay graph is shown in Figure \ref{fig:PVD}-F. All its edges are contained either in the k-NN graph (Figure \ref{fig:PVD}-A) or in the edges coming
    from cell intersections (Figure \ref{fig:PVD}-D).
\end{itemize}

The Delaunay triangulation, if needed, is deduced from the combinatorics of the Voronoi diagram. In 3D, a tetrahedron $(i,j,k,l)$ is generated each time the intersection between the dominance regions $\Pi^+(i,j)$, $\Pi^+(i,k)$,
$\Pi^+(i,l)$ define a vertex of $\Vor_i$. To avoid obtaining the tetrahedra four times, one can keep only the one for which $i$ is the smallest index.
To make sure the same tetrahedra are generated from different Voronoi cell, one needs to take care of configurations with cospherical points, that generate Voronoi vertices of degree potentially larger than 4 \textcolor{black}{(that is, incident to more than 4 Voronoi regions)}, hence several possible tetrahedralizations. A simple solution is to use the \textcolor{black}{so-called} classical symbolic perturbation approach \cite{10.1145/77635.77639}. \textcolor{black}{Computing a Voronoi diagram requires one to deduce some combinatorics from the relative positions of geometric objects, classifying them as 'above' or 'below' for instance. This is often expressed as the sign of polynomial function of the point's coordinates called a predicate. The configurations where the polynomial is zero - for instance co-spherical points in Voronoi diagrams - require a special treatment. One possibility is to \emph{symbolically} apply a permutation to each point's coordinate, and use the sign of the first non-zero term of the Taylor expansion.}
The only predicate used by the computation of the cells $V^\E_i$ requires to classify a Voronoi vertex, that is, the intersection of three bisectors, relative to a fourth bisector, which is equivalent to the {\tt in\_sphere} predicate (for Voronoi diagrams), or its weighted generalization (for Laguerre diagrams). Symbolically-perturbed implementations of these predicates are readily available \cite{shewchuk97a,DBLP:journals/cad/Levy16,WEB:GEOGRAM,cgal:eb-23b}.

This algorithm is very similar to \emph{star splaying} \cite{10.1145/1064092.1064129},
that also starts from a subset of the Delaunay graph, and finds the missing edges. In other words, one may describe this algorithm as ``star splaying seen from the point of view of the Voronoi cells". It adds an alternative way of considering the neighborhoods, in addition to the three points of view usually associated with star splaying (``star of Delaunay vertex", ``convex hull of ray" and ``cross-section of cone").
The four points of view are formally equivalent, but focusing on Voronoi cells has three practical consequences:
\begin{itemize}
    \item it may be a question of taste, but one may find the explanation in terms of
    Voronoi cells shown in Figure \ref{fig:PVD} easier to follows: simply put, the dual of ``star splaying" is ``Voronoi clipping" or ``Voronoi trimming";
    \item instead of storing all the vertice's neighborhoods, one can store only the graph, and recompute the Voronoi cells as need be. While this has an additional computational cost, it saves substantial amounts of memory;
    \item by considering the Voronoi cells explicitly, one can exploit their relations with the boundary of the domain to optimize point exchanges in the distributed algorithm, as explained in the next section.
\end{itemize}


\subsection{DVD: Distributed Voronoi Diagram}
\label{sec:DVD}

We now suppose that the domain is partitioned into $M$ regions $(R_k)_{k=1}^M$, and that the pointset $\bX$ is partitioned accordingly, into $M$ subsets
$\bX_k = \bX \cap R_k$. In addition, we suppose that each region $R_k$ is represented and processed by a different node of a cluster. Each node $k$ is going to exchange some points with other nodes in order to ensure that it obtains all the necessary information to compute all the Voronoi cells of the points $\bX_k$.

\begin{algorithm}
    By-region parallel Voronoi Diagram
$$
    \begin{array}{l}
       \begin{array}{ll}
          \mbox{\bf Input:} & \mbox{\rm the regions } \left\{R_k\right\}_{k=1}^M \mbox{\rm and the pointsets } \left\{\bX_k\right\}_{k=1}^M \\[2mm]
          \mbox{\bf Output:} &\ M\ \mbox{\rm graphs } \E_k \mbox{\rm, such that}\ \Vor_i = \V^{\E_k}_i\ \forall i\ \mbox{\rm such that}\ \bx_i \in R_k  \\[3mm]
       \end{array}\\[5mm]
       \hline\\[5mm]
       \begin{array}{ll}
         (1) & \mbox{\bf for } k = 1 \ldots M, \quad \E_k \leftarrow \Del(\bX_k) \\[2mm]
         (2) & \mbox{\bf for } k = 1 \ldots M, \quad \bY_k \leftarrow \left\{ \bx_i \in R_l \ |\ l \neq k\ \mbox{\rm and } \V_i^{\E_l} \cap R_k \neq \emptyset \right\} \\[3mm]
         (3) & \mbox{\bf for } k = 1 \ldots M, \quad \E_k \leftarrow \Del(\bX_k \cup \bY_k) \\[2mm]
         (4) & \mbox{\bf for } k = 1 \ldots M, \quad \bZ_k \leftarrow \left\{ \bx_j \ |\ \exists l \neq k, \exists \hedge{i}{j} \in \E_l, \bx_i \in \bX_k, \bx_j \in \bX_l \right\} \\[2mm]
         (5) & \mbox{\bf for } k = 1 \ldots M, \quad \E_k \leftarrow \Del(\bX_k \cup \bY_k \cup \bZ_k)
       \end{array}
    \end{array}
$$
\label{algo:DVD1}
\end{algorithm}

As in the previous section, in step (1), the algorithm starts by computing a graph $\E$, but this time the graph is distributed over a set of $M$ nodes, $\E = \E_1 \cup \ldots \cup \E_M$, and each component $\E_k$ stored in a node corresponds to the Delaunay graph $\Del(\bX_k)$ of the vertices of $\bX$ that fall in the region $R_k$ associated with that node. To compute the Delaunay graph, one may use the standard Bowyer-Watson algorithm \cite{DBLP:journals/cj/Bowyer81,journals/cj/Watson81}, or Algorithm \ref{algo:PVD} in the previous section. 
Once the Delaunay graph $\E_k$ is computed, most cells $\V^{\E_k}_i$
correspond to the Voronoi cells $\Vor_i$ (since $\E_k$ is the Delaunay graph of $\bX_k$), except for some cells on the boundary, that are larger than the Voronoi cells (Observation \ref{obs:cells}). The edges $\edge{i}{j}$ that are missing in $\E_k$ correspond to pairs of cells with non-empty intersections (Observation \ref{obs:subgraph}), stored in different nodes.
To have in each $\E_k$ all the edges of the Delaunay graph, we need to determine both the relation $i$ \emph{is clipped by} $j$
(that corresponds to the oriented edge $\hedge{i}{j}$ and the relation $i$ \emph{is clipping} $j$ (that corresponds to the oriented edge $\hedge{j}{i}$), hence we have an algorithm that operates in two phases (steps (2,3) and steps (4,5)): in step (2), each node $k$ gathers the points $\bx_i \in R_l$ with a cell $\V^{E_l}_i$ that has a non-empty intersection with $R_k$ (to detect the missed \emph{is clipped by} relations). In step (3), the graphs are updated. In step (4), the so-discovered new edges are symmetrized, by sending back the neighbors (\emph{is clipping} relations). Each graph $\E_k$ obtained in the node $k$ at step (5) has all the edges to properly compute the Voronoi cells of the points in $R_k$.

\begin{figure}
    \centering
    \includegraphics[width=\textwidth]{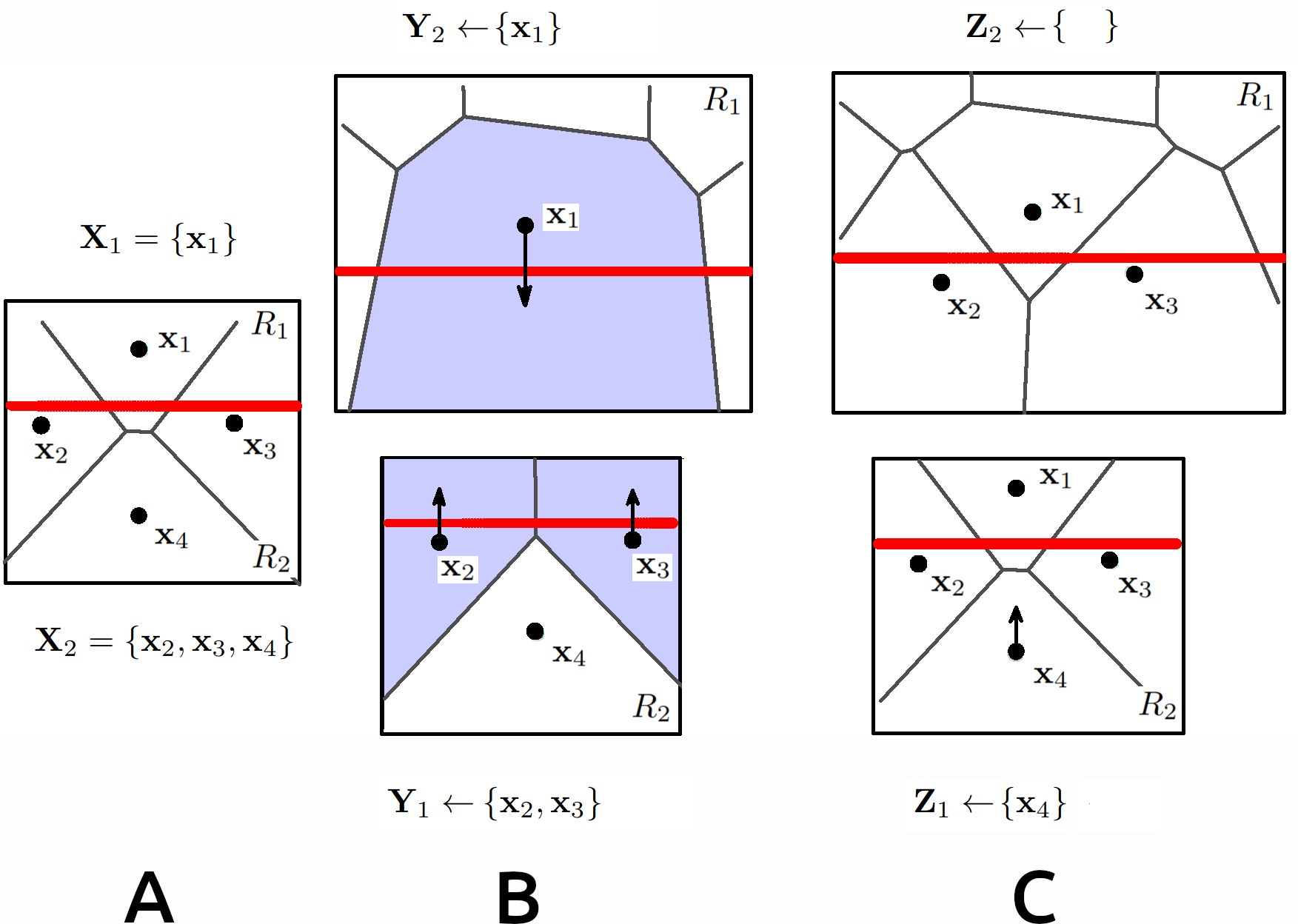}
    \caption{A simple example showing why two steps of point exchange are needed in the distributed Voronoi diagram algorithm.}
    \label{fig:DVD_example}
\end{figure}

\paragraph*{\bf Why do we need the two steps (2) and (4) ?} Our goal is to represent,
in a distributed manner, a graph $\E = \bigcup_k \E_k$ that has all the edges of the Delaunay
graph $\Del(\bX)$. For each (unoriented) edge $\edge{i}{j} \in \Del(\bX)$, we need to find the two
oriented edges $\hedge{i}{j}$ (that corresponds to the relation $i$ \emph{is clipped by} $j$) and
$\hedge{j}{i}$ (that corresponds to the relation $i$ \emph{is clipping} $j$) in all the nodes that contain $\bx_i$ or $\bx_j$, hence two steps are needed. Let us examine them more closely:

\begin{itemize}
\item Step (2) ensures that all for all $\bx_i$ in a region $R_k$, all the edges $\hedge{j}{i}$ such that $\edge{i}{j} \in \Del(\bX)$ are in the $\E_{l,l\neq k}$ (that is, the $i$ \emph{is clipping} $j$ relation, for the $x_j$'s in the other regions);
\item Step (4) ensures that for all $\bx_i$ in a region $R_k$, all the edges $\hedge{i}{j}$ such that
$\edge{i}{j} \in \Del(\bX)$ are in $\E_k$ (that is, the $i$ \emph{is clipped by} $j$ relation, for the $x_j$'s in the other regions).
\end{itemize}
Hence step (4) is simply a combinatorial operation that \emph{symmetrizes} the oriented edges discovered at step (2).  \\

{\textcolor{black}{Note that observations \ref{obs:cells}, \ref{obs:subgraph} hold in arbitrary dimension, as well as algorithm \ref{algo:PVD}. To help grasping an intuition of what the algorithm does, we shall now examine the effect effect of the two steps on a particular 2D example, illustrated in Figure \ref{fig:DVD_example}}. One wants to compute the Voronoi diagram shown on the left, using two regions $R_1$ and $R_2$. The boundary between $R_1$ and $R_2$ is shown in red. The sets of points $\bX_1$ and $\bX_2$ are initialized with the points contained by regions $R_1$ and $R_2$ respectively, shown in  Fig. \ref{fig:DVD_example}-A. In the first step (Fig. \ref{fig:DVD_example}-B),
\begin{itemize}
\item Region $R_1$ contains $\bx_1$. The cell $\V^\E_1$ of $\bx_1$ has an intersection with the boundary, hence $\bx_1$ is sent to $R_2$.
\item Region $R_2$ contains $\bx_2, \bx_3$ and $\bx_4$. The cells of $\bx_2$ and $\bx_3$ have an intersection with the boundary, and are sent to $R_1$.
\end{itemize}
In the second step (Fig. \ref{fig:DVD_example}-C), region $R_1$ knows $\bx_1$, $\bx_2$ and $\bx_3$. The cell $\V^\E_1$ is not correct, because it does not know (yet) its neighbor $\bx_4$, but the neighboring relation $\edge{1}{4}$ is known by $R_2$, that sends $\bx_4$ to $R_1$.

With this by-region granularity, Algorithm \ref{algo:DVD1} can be implemented in a cluster. All the variables with index $k$ are stored
in a specific node of the cluster.  The five loops in steps
(1)-(5) are executed in parallel, by all the nodes in the cluster. The only inter-node communications are at steps (2) and (4) (plus the initial
broadcast and partition of $\bX$ into the $R_k$'s). In algorithm \ref{algo:DVD1}, at steps (2) and (4) a node $k$ \emph{gathers} points from potentially all the
other nodes. From an implementation point of view, it is more natural to have the nodes interconnected with only the neighboring nodes through communication channels. In this setting, the node that corresponds to region $R_k$ is connected to the nodes of the adjacent regions. During the execution of
the algorithm, each node \emph{sends} the points where they are needed. This alternative point of view leads to the following algorithm, executed by each node in parallel (the loops on the nodes $k$ are omitted):

\begin{algorithm}
    Distributed Voronoi Diagram (algorithm for a node $k$)
$$
    \begin{array}{l}
       \begin{array}{ll}
          \mbox{\bf Input:} & \mbox{\rm the regions } \left\{R_k\right\}_{k=1}^M \mbox{\rm and the pointsets } \left\{\bX_k\right\}_{k=1}^M \\[2mm]
          \mbox{\bf Output:} &\ M\ \mbox{\rm graphs } \E_k \mbox{\rm, such that}\ \Vor_i = \V^{\E_k}_i\ \forall \bx_i \in R_k  \\[3mm]
       \end{array}\\[5mm]
       \hline\\
       \begin{array}{ll}
         (1)  &  \E_k \leftarrow \Del(\bX_k) \\
         (2)  & \mbox{\bf for each } \bx_i \in \bX_k \\
         (3)  & \quad \mbox{\bf for each } R_l\ \mbox{\rm neighbor of } R_k \\
         (4)  & \quad \quad \mbox{\bf if } \V^{\E_k}_i \cap R_l \neq \emptyset\ \mbox{\bf then } \mbox{\rm send } \bx_i\ \mbox{\rm to neighbor node } l \\
         (5)  & \bY_k \leftarrow \mbox{\rm receive points from neighbor nodes} \\
         (6)  & \mbox{\bf sync} \\
         (7)  & \E_k \leftarrow \Del(\bX_k \cup \bY_k) \\
         (8)  & \mbox{\bf for each } \hedge{i}{j} \in \E \\
         (9)  & \quad l \leftarrow \mbox{Region}(\bx_i) \quad ; \quad m \leftarrow \mbox{Region}(\bx_j) \\
         (10) & \quad \mbox{\bf if } l = k\ \mbox{\bf and } m \neq k\ \mbox{\bf then } \mbox{\rm send } \bx_j\ \mbox{\rm to neighbor node } m  \\
         (12) & \bZ_k \leftarrow \mbox{\rm receive points from neighbor nodes} \\
         (13) & \mbox{\bf sync} \\
         (14) & \E_k \leftarrow \Del(\bX_k \cup \bY_k \cup \bZ_k)
       \end{array}
    \end{array}
$$
\label{algo:DVD2}
\end{algorithm}
where \textbf{sync} is a synchronization point for all the threads (one waits there until everybody has reached this point), and where $Region(l)$, $Region(m)$ denote the region that
sent $l$ (resp. $m$) at step (5) (each node keeps track of who sent each point).

\textcolor{black}{The algorithm detailed above, running on each cluster node $k$, sends a vertex to another node $l$ each time the Voronoi cell of the vertex has an intersection with the region $R_l$ associated with the node $l$, which is straightforward to determine if the set of regions form a regular grid. In our specific case (cosmological simulation), each region contains a large number of points filling the
  entire region (even if density can vary a lot locally), see Figure \ref{fig:laguerre_100M}. While the situation where a Voronoi cell traverses several regions may theoretically happen in general, in our specific application, assuming that it does not happen is a viable assumption. Hence it is possible to make an optimization, simplifying the connection graph between the nodes of the cluster:}
Each cluster node $k$, that represents region $R_k$ is typically connected to nodes corresponding to the regions adjacent to $R_k$.
If regions are bounded by polyhedra, adjacent regions can have a face, an edge or a vertex in common with $R_k$. Typically, one can
use a regular grid, with up to four regions incident to each edge and up to eight regions incident to each vertex. If a candidate Voronoi cell encroaches a facet, then the corresponding point is sent to the region connected to the other
side of the facet. If the candidate Voronoi cell encroaches an edge or a vertex, then the point is sent to all the regions incident to
the edge or the vertex. Note that \textcolor{black}{modified with this optimization}, Algorithm \ref{algo:DVD2} does not work if a cell traverses a region completely. If one wants to support this condition, \textcolor{black}{besides not doing the optimization / using the unmodified algorithm}, one can also execute the two points exchange steps of the algorithm \textcolor{black}{on the reduced cluster node graph} until each point was sent to all the regions it has a non-empty intersection with.

The algorithm also works for Laguerre diagrams, but one needs to take care: in a Laguerrre diagram,
the cell of $\bx_i$ does not necessarily contain $\bx_i$, and can even have no intersection with $R_k$ (it can be totally on the other side, inside $R_l$).

This algorithm is similar to the one in \cite{7013068}, that shares the same global structure (and the same restriction that a cell should not traverse a region), with the two steps to recover the missing edges in the Delaunay graph. The main difference is in Step (4): instead of considering the cells, their method is based on the vertices, that should satisfy the empty ball property. The vertex common to cells $\V^\E_i, \V^\E_j, \V^\E_k$ is a Voronoi vertex if the circumscribed ball
to the triangle $(i,j,k)$ does not contain any point of $\bx_i$. Hence, the points $\bx_i,\bx_j,\bx_k$
are sent to the regions that have a non-empty intersection with the ball (in 3D, there are 4 cells meeting at a vertex, and it is the circumscribed ball to a tetrahedron). Infinite cells require a special treatment, and consider the cone defined by the edges (triangles in 3D) on the border of the Delaunay graph.
On the one hand, computations are simpler than in our algorithm that needs computing Voronoi cells explicitly. On the other hand, the objects considered for sending a point $\bx_i$ to the neighboring cells are larger than in our case.
In our case, the cost of sending more point than necessary is larger than the cost of computing the cells. Moreover, in our cosmological application shown in the next section, we can have a huge variation of pointset density, potentially creating pencil-shaped Voronoi
cells on the border. Hence it is important to be able to determine which elements of the region boundary they encroach in order to avoid
sending the point to too many nodes.

\section{Large-scale optimal transport for cosmology}
\label{sec:OT}

\begin{figure}
    \colorbox{black}{
    \begin{minipage}{\textwidth}
    \centerline{
       \includegraphics[width=0.35\textwidth]{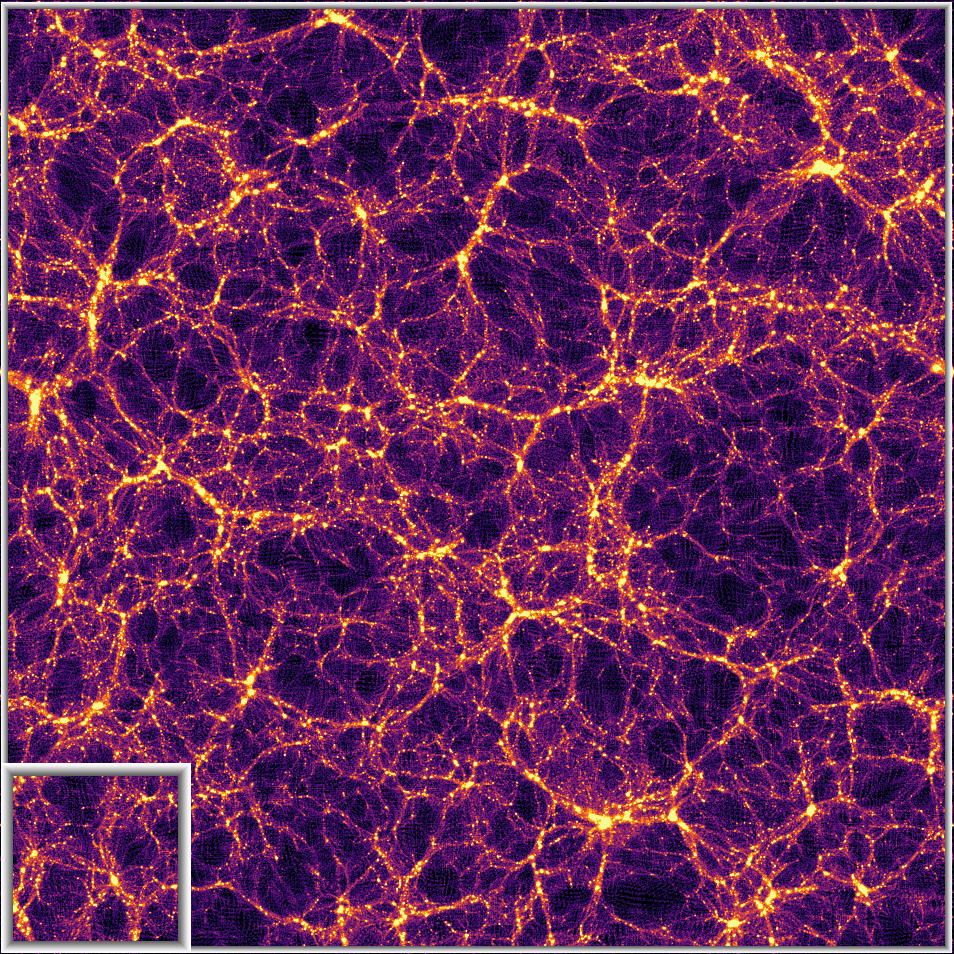}
       \hspace{23mm}
       \includegraphics[width=0.35\textwidth]{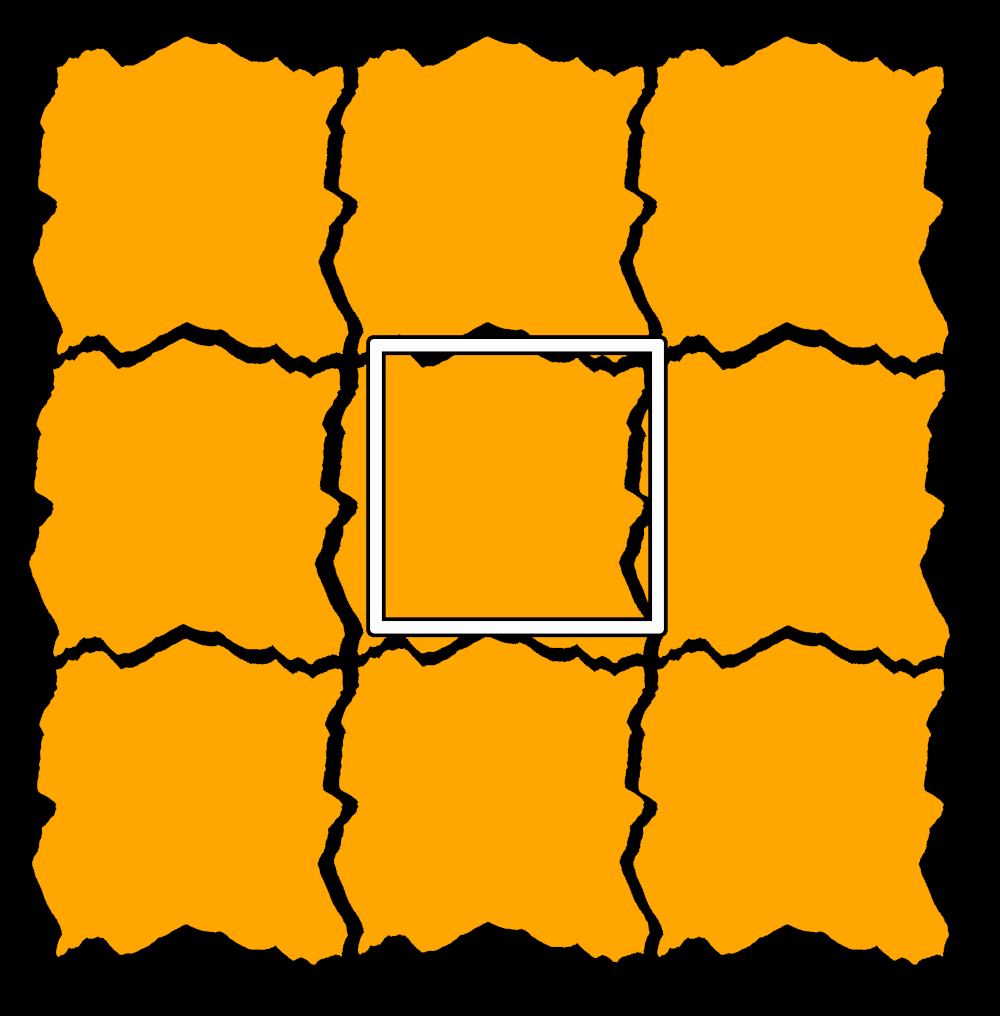}

    }
    \centerline{\color{white}A \hspace{0.5\textwidth} B}
    \centerline{
     \includegraphics[width=\textwidth]{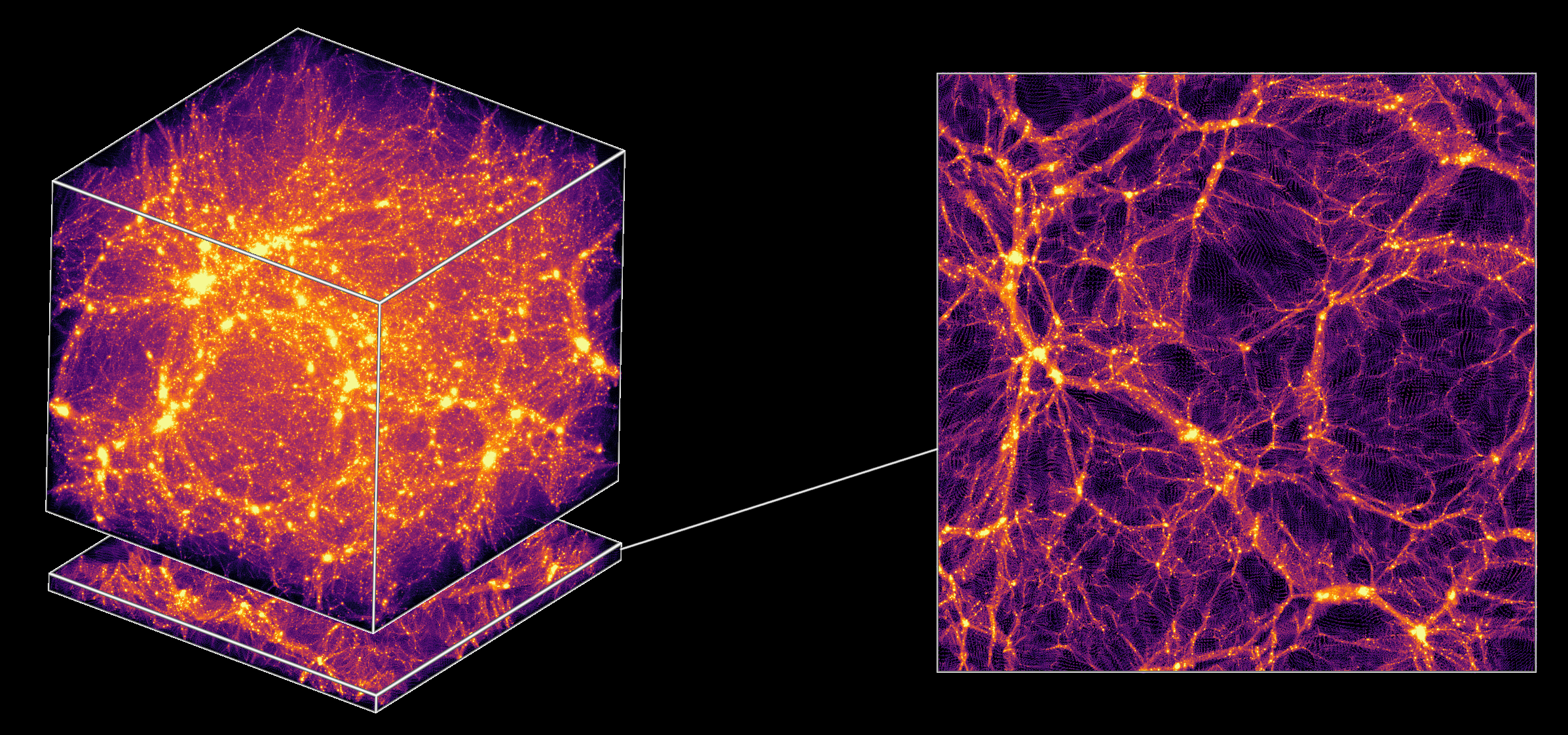}
    }
    \centerline{\color{white}C}
    \centerline{
    \includegraphics[width=\textwidth]{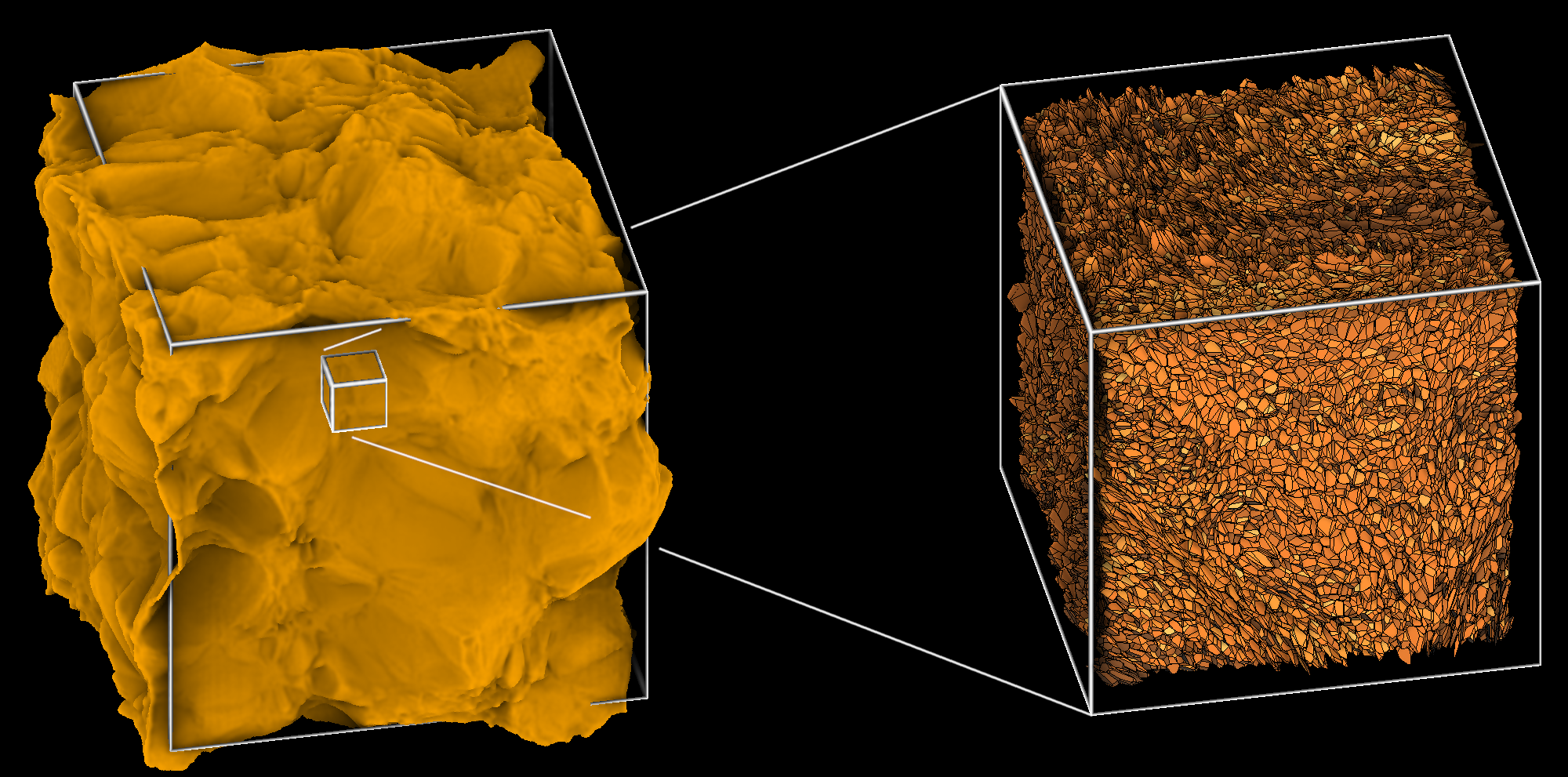}
    }
    \centerline{\color{white}D}
    \end{minipage}}
    \caption{A: Thin-slice in a cosmological simulation of a 300 Mpc/h cube. Ultimately, the goal will be to compute optimal transport with 10 billion points. The cube will be decomposed into $5 \times 5 \times 5$ subvolumes (one of them highlighted). For now, the algorithm is tested and analyzed with periodic boundary conditions (B) in one of the subvolumes (C) with 100 million points. The reconstructed initial condition (D).
    }
    \label{fig:laguerre_100M}
\end{figure}

\begin{figure}
    \centering
    \includegraphics[width=\textwidth]{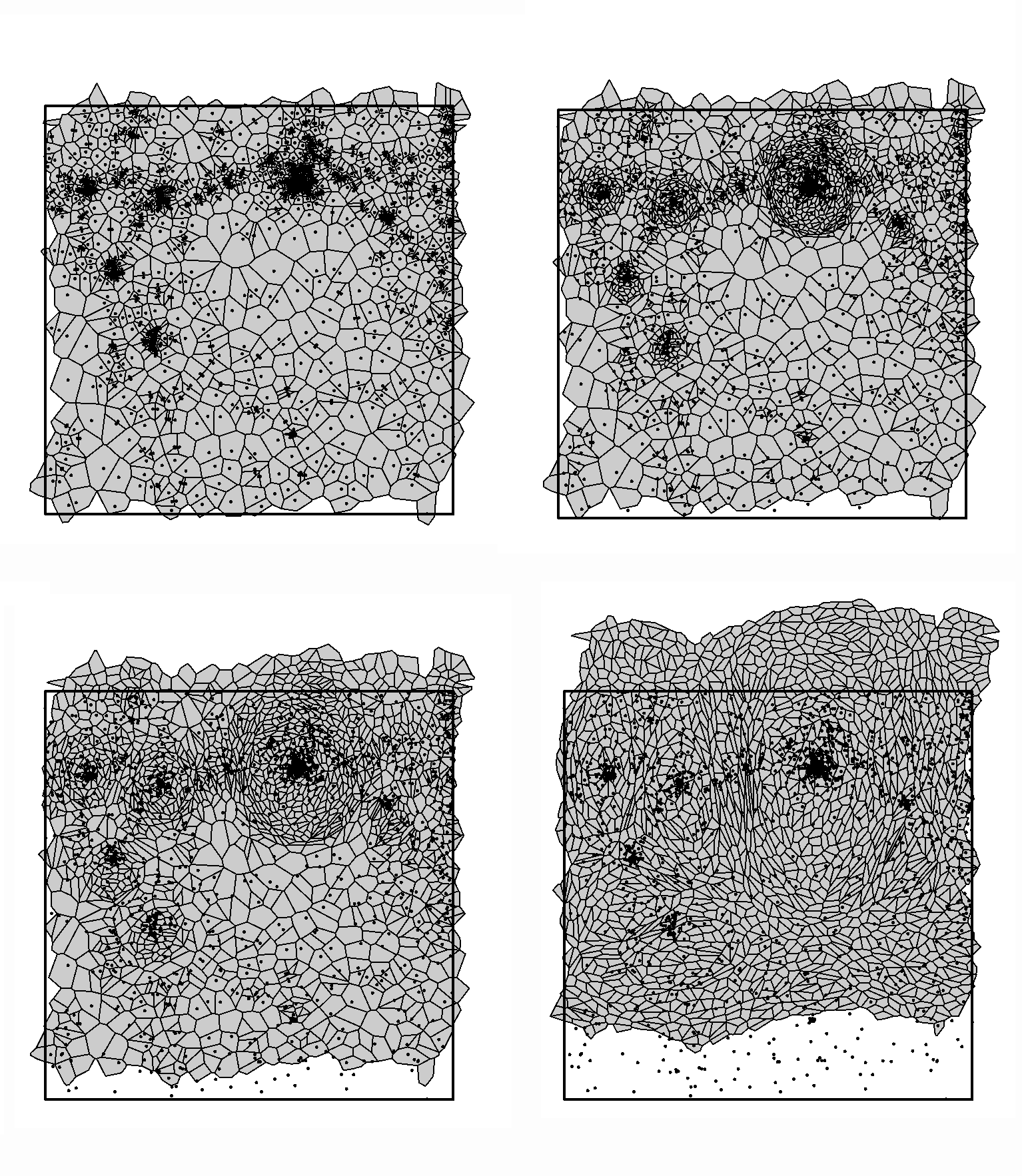}
    \caption{Newton iterations of semi-discrete optimal transport with periodic boundary conditions}
    \label{fig:OT_iter_periodic}
\end{figure}

\subsection{An example in early Universe reconstruction}

We shall now see how the algorithm behaves in a practical scenario, in Early Universe Reconsruction. Figure \ref{fig:laguerre_100M}-A shows the result of a cosmological simulation of structure formation,
in a cube of 300 Mpc/h edge length in cosmological units\footnote{300 Mpc/h stands for 300 mega parsecs \textcolor{black}{(Mpc) in co-moving coordinates (/h). Co-moving coordinates factor-out the expansion of the Universe}. From this 300 Mpc/h threshold, the large-scale structure of the Universe starts to be homogeneous. For reference, the nearest star is 1.399 parsecs away from the sun, and the radius of the observable Universe is around 14.25 giga parsecs.}. The figure shows a thin slice (5 Mpc/h thick) of the cube. Each point corresponds to a galaxy cluster. The goal of cosmological reconstruction \cite{EURNature,EUR,10.1093/mnras/stab1676,vhauss_prl_2022,nikak_prl_2022} is to trace the trajectories of these galaxy clusters back in time, supposing that they started from a uniform initial condition. Under
some simplifying assumptions (convexity of the integrated potential), it can be restated as a semi-discrete optimal transport problem. The goal, in the future, is to apply such optimal-transport-based reconstruction to massive pointsets, with billions of particles. When reconstructing from observational data, such gigantic-scale problems appear when modeling the unobserved dark matter as additional particles \cite{nikak_prl_2022}. They are also needed in cosmological simulation of modified gravity \cite{levy2024monge}. Typically, for our cube of 300 Mpc/h, it would be interesting to be able to solve optimal-transport problems with $2560^3$ points, which would give sufficient resolution to track fine-scaled dynamics. The idea will be to run the Distributed Voronoi Diagram on a PC cluster with 125 machines, each machine will be responsible for a cube of 60 Mpc/h with 100 million particles ($512^3$), depicted as a small square
in the corner of Figure \ref{fig:laguerre_100M}-A. Such a large-scale experiment will be the subject of later work. For now, we conducted a "proof-of-concept" experiment:
as often done in cosmology and material science, to simulate a large volume of a globally homogeneous medium, one can use periodic boundary conditions. For instance, one could imagine considering the small cube in Figure \ref{fig:laguerre_100M} and connect its opposite faces before reconstructing the motion of the galaxy clusters. One then obtains an initial condition that paves the space, as the 2D example shown in
Figure \ref{fig:laguerre_100M}-B.

Figure \ref{fig:laguerre_100M}-C
shows the final timestep of a cosmological simulation, with 100 million points, in a cube of 60 Mpc/h edge length (same size of the small cube in Figure \ref{fig:laguerre_100M}-A) with a periodic boundary condition. We made this simulation with a slightly modified version of \cite{Couchman1994HydraAA} (modifications are to make it work with $512^3$ points, original version was limited to $256^3$).
Starting from this pointset, one can reconstruct the initial condition using optimal transport. In the end, one obtains a reconstructed initial condition, \textcolor{black}{the union of the Laguerre cells shown as the yellow shape. This shape paves the space, its curvy boundaries correspond to matter that crossed the periodic boundaries.} It represented by a Laguerre diagram (Figure \ref{fig:laguerre_100M}-D), and each Laguerre cells correspond to the matter that lumped to form one of the galaxy clusters.
The closeup in Figure \ref{fig:laguerre_100M}-D shows what these Laguerre cells look like in the central region of the cube.

Computing Voronoi and Laguerre diagrams with periodic boundary conditions is just a particular case of the algorithm presented in this article. In a 2D version, one considers 9 regions, indexed by ``translation vectors" with
coefficients in $\{-1,0,1\}$:

\begin{center}
\begin{tabular}{c|c|c}
  $R_{-1,1}$ & $R_{0,1}$ & $R_{1,1}$ \\
  \hline
  $R_{-1,0}$ & $R_{0,0}$ & $R_{1,0}$ \\
  \hline
  $R_{-1,-1}$ & $R_{0,-1}$ & $R_{1,-1}$
\end{tabular}
\end{center}

\begin{table}[]
\color{black}
    \centerline{
    \begin{tabular}{c|c|c|c|c|c|c|c|c}
         Iter.  & \multicolumn{4}{c|}{phase I} & \multicolumn{3}{c|}{phase II} & total \\
                & npts cross & npts out & $t_{class}$ & $t_{insert}$ & npts & $t_{class}$ & $t_{insert}$  & $t$   \\
        \hline
           0 & 646800 & 0 & 2.2 & 0.5 & 233050 & 3.4 & 0.3 & 22.2 \\
1 & 646894 & 1 & 2.1 & 0.5 & 233042 & 3.2 & 0.3 & 18.2 \\
2 & 648153 & 455 & 2.0 & 0.5 & 232706 & 3.2 & 0.3 & 18.1 \\
3 & 650365 & 2577 & 2.2 & 0.6 & 232284 & 3.2 & 0.4 & 19.2 \\
4 & 656861 & 13925 & 2.0 & 0.6 & 232124 & 3.2 & 0.3 & 17.7 \\
5 & 665621 & 33767 & 2.1 & 0.6 & 232375 & 3.2 & 0.3 & 18.3 \\
6 & 679112 & 74305 & 2.0 & 0.6 & 233205 & 3.4 & 0.3 & 19.6 \\
7 & 698330 & 154415 & 2.2 & 0.6 & 234937 & 3.1 & 0.3 & 18.8 \\
8 & 727751 & 293985 & 1.6 & 0.7 & 237624 & 3.2 & 0.3 & 17.8 \\
9 & 769940 & 518739 & 2.0 & 0.9 & 243229 & 3.2 & 0.4 & 22.0 \\
10 & 773413 & 540357 & 1.9 & 0.9 & 243777 & 3.4 & 0.3 & 18.9 \\
11 & 780948 & 583207 & 2.2 & 0.8 & 245062 & 3.1 & 0.4 & 19.0 \\
12 & 787834 & 624921 & 2.1 & 0.8 & 246301 & 3.4 & 0.3 & 19.7 \\
13 & 838950 & 984704 & 2.1 & 1.0 & 257811 & 3.4 & 0.3 & 22.2 \\
14 & 874334 & 1306647 & 2.3 & 1.2 & 269045 & 3.5 & 0.4 & 21.9 \\
15 & 921254 & 1898587 & 2.0 & 1.3 & 289628 & 3.5 & 0.4 & 18.7 \\
16 & 952246 & 2407000 & 1.9 & 1.5 & 306489 & 3.9 & 0.5 & 21.5 \\
17 & 978500 & 2873474 & 2.0 & 1.6 & 320861 & 4.1 & 0.4 & 19.8 \\
18 & 1024563 & 3768530 & 2.2 & 1.5 & 344896 & 4.5 & 0.4 & 20.2 \\
19 & 1057950 & 4580610 & 2.0 & 1.8 & 363521 & 4.7 & 0.4 & 21.1 \\
20 & 1129350 & 6117650 & 2.3 & 2.6 & 394322 & 4.7 & 0.5 & 22.8 \\
21 & 1188534 & 7459069 & 2.1 & 2.7 & 417273 & 5.2 & 0.5 & 23.9 \\
22 & 1234116 & 8635647 & 2.5 & 3.1 & 435938 & 5.4 & 0.5 & 24.4 \\
23 & 1266707 & 9666107 & 2.3 & 3.5 & 451221 & 5.5 & 0.6 & 24.5 \\
24 & 1322035 & 11457655 & 2.3 & 3.8 & 478621 & 5.8 & 0.6 & 29.5 \\
25 & 1359821 & 12802087 & 2.2 & 4.3 & 499038 & 5.8 & 0.6 & 25.5 \\
26 & 1411051 & 14836994 & 2.3 & 5.8 & 532127 & 6.3 & 0.3 & 27.3 \\
27 & 1411345 & 14853396 & 2.5 & 4.9 & 532418 & 6.3 & 0.3 & 27.1 \\
28 & 1413260 & 14984207 & 2.7 & 4.9 & 534741 & 6.7 & 0.3 & 31.1 \\
29 & 1421313 & 15472725 & 2.6 & 5.3 & 542958 & 6.6 & 0.3 & 32.1 \\
30 & 1432308 & 16207255 & 2.4 & 6.3 & 557217 & 6.6 & 0.4 & 28.8 \\
31 & 1443959 & 16951863 & 2.7 & 5.7 & 574707 & 6.8 & 0.3 & 33.2 \\
32 & 1444741 & 16962135 & 2.5 & 5.8 & 575423 & 6.9 & 0.4 & 32.9 \\
33 & 1444765 & 16962171 & 2.4 & 5.3 & 575465 & 7.0 & 0.4 & 32.1 \\
34 & 1444765 & 16962171 & 2.4 & 5.4 & 575466 & 6.8 & 0.4 & 32.6 \\
\hline
    \end{tabular}
    }
    \color{black}
    \caption{Statistics of the Laguerre diagram computations during the Newton iterations of a semi-discrete optimal transport problem with 134M points and periodic boundary
    conditions. All times are in s.}
    \label{tab:EUR}
\end{table}

\begin{table}[]
    \color{black}
    \centering
    \begin{tabular}{l|c|c}
         \hline
         total time &  100\% & 1:12:37 \\
         Laguerre & 29.8\% &  0:20:59 \\
         Linear solve & 37.89\% & 0:27:31 \\
         Eval gradient & 1.3\% & 58s \\
         Assemble Hessian & 28.1\% & 0:20:29 \\
         Misc & 3.65\% & 0:02:39 \\
         \hline
    \end{tabular}
    \color{black}
    \caption{Optimal transport for 134M points with periodic boundary conditions. Timing breakdown}
    \label{tab:timing_breakdown}
\end{table}

\begin{figure}
    \centering
    \includegraphics[width=0.75\textwidth]{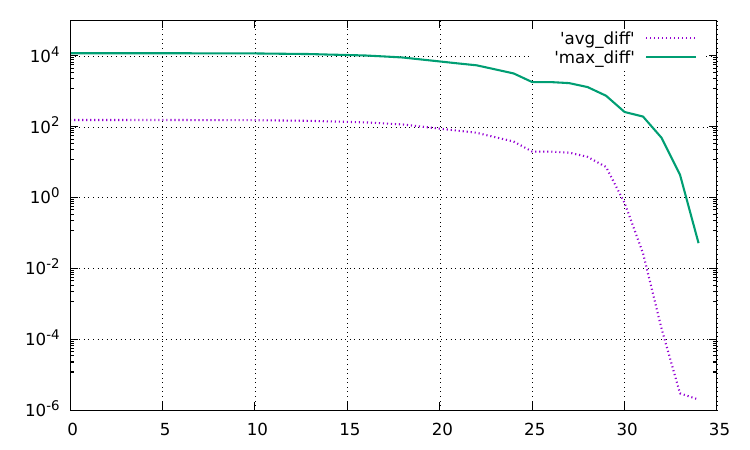}
    \includegraphics[width=0.75\textwidth]{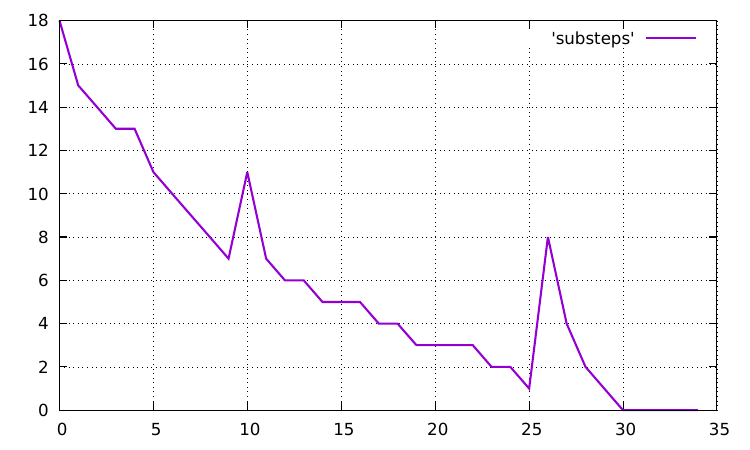}
    \caption{Top panel: Average cell difference and worst cell difference in function of Newton iterations (in percent of target cell measure, logarithmic scale). Bottom panel: number of substeps in each Newton iteration.}
    \label{fig:diff}
\end{figure}

In practice, only $R_{0,0}$ is stored. In steps (4) and (10) of Algorithm \ref{algo:DVD2}, it ``sends points to itself", by creating new copies
of the points, translated according to which facet/edge/corner of the unit
cube was traversed.

The example shown in Figure \ref{fig:laguerre_100M}-C,D is a good ``stress test" for semi-discrete optimal transport. Besides the number of points that starts to be large, and the additional difficulty of the periodic boundary conditions \cite{BOURNE2023104023}, this optimal transport problem is especially challenging, because at this small scale (relative to the size of the visible Universe), the variations of density are considerable: distances between the points vary by 5 orders of magnitude. Some regions are very dense, and some regions are nearly empty. This has several consequences for the iterative algorithm that computes the Laguerre diagram. As shown in the 2D example in Figure \ref{fig:OT_iter_periodic}, it starts with a Voronoi diagram, then evolves the weights of a Laguerre diagram to find the (unique) Laguerre diagram such that the areas of the Laguerre cells match the masses of the points. As can be seen, the large variations of points density makes the first iterations quite ``explosive". Then, as one can see, the Laguerre cells ``travel" in space, to reach the position where the matter that corresponds to each cluster started from. As one can see, a Laguerre cell does not necessarily contain its generating point (unlike Voronoi cells).
Another consequence of the high variations of density is that the matter that lumped to form the structures traveled a lot (up to 15\% of cube edge length). \textcolor{black}{In figure \ref{fig:laguerre_100M}-D, they correspond to the parts of the yellow shape that crossed the boundary of the cube}. In terms of our algorithm, it means that a significant number of points will be sent through the boundaries of the domain. The same behavior is expected when we will be running the algorithm on a cluster grid, with different pointsets in each cube.

In Table \ref{tab:EUR}, we report different timings and statistics on
Algorithm \ref{algo:DVD2} executed in the context of semi-discrete optimal transport for early Universe reconstruction, with the pointset shown in
Figure \ref{fig:laguerre_100M}-B. The algorithm converged after 34 Newton iterations, which took \textcolor{black}{1h 15min in total,
on a dual-CPU AMD EPYC 9754 (256 cores per CPU)}.
For each iteration, the table reports the number of points with a Laguerre cell that crosses the boundary of the domain (npts cross), the number of points completely on the other side of the boundary (npts out), the time to classify the points ($t_{class}$), to
insert the points into the triangulation ($t_{insert}$), for both phases
of the algorithm (steps (1) to (6) and (7) to (14) in Algorithm \ref{algo:DVD2} p. \pageref{algo:DVD2}), as well as the total time for computing the Laguerre diagram
(this includes the time to compute the initial triangulation in $R_{0,0,0}$). As can be seen, each iteration takes longer and longer, which is because the Laguerre cells travel a longer and longer distance
(like in Figure \ref{fig:OT_iter_periodic}), and a larger and larger number of points needs to be sent. Phase II takes a shorter time as compared to Phase I, because it is a purely combinatorial operation, whereas Phase I needs to compute the Laguerre cells and their relations with the boundary. However, execution time remains reasonable, even in the last iterations, that need to send more than 10\% of the points.
Figure \ref{fig:diff} shows the convergence and the number of KMT substeps used by each Newton iteration. The average difference between a cell's measure $|\mbox{Lag}^\psi_i|$ and the target measure $\nu_i$ (expressed in \% of target cell measure), as well as the maximum difference in function of the iteration are both displayed.
As can be seen, the first Newton iterations are challenged by the huge difference of density, and use a tiny $\alpha = 2^{-18}$ descent parameter. Starting from the $30^{th}$ iteration, the algorithm enters a regime with quadratic speed of convergence, and the measure error of the worst cell quickly decreases. \textcolor{black}{Table \ref{tab:timing_breakdown} shows the timing breakdown of the optimal transport computation. The most time consuming phases are the linear solves, the computation of the Laguerre diagrams and matrix assemblies.}

\section{Conclusions and future works}
The numerical experiments presented here are a ``proof of concept": this ``mock-up" numerical experiment demonstrates the feasibility of the approach for\ \ the\ \ large-scale experiment illustrated in Figure\ \ \ref{fig:laguerre_100M}-A:\ \ \ the distributed \\
\noindent Voronoi diagram algorithm exchanges a larger and larger number of points as Laguerre cells travel through the boundaries of the periodic domain. In the large-scale scenario, a similar number of points will be exchanged (but this time, with neighboring cluster nodes instead of the same cube seen from periodic boundaries).
The experiment also demonstrates that the algorithm can handle the huge variations of density, that challenge both its geometric part (Laguerre diagram) and numeric part (linear systems).

The software prototype is useful as it is for optimal-transport reconstruction with periodic boundary conditions, and was successfully applied it to problems with sizes ranging from 16M to 300M points \cite{10.1093/mnras/stab1676,vhauss_prl_2022,nikak_prl_2022,PhysRevD.108.083534}. Now the goal is to scale-up to problems of 1 billion points and above. This will make it possible to work with a volume of 300 Mpc/h (Figure \ref{fig:laguerre_100M}-A) with the same resolution, and better measure different phenomena at a fine scale. This means running a cube of $5 \times 5 \times 5$ interconnected instances of the 60 Mpc/h reconstruction shown here on a cluster, which will be done in future works. Besides the implementation of the DVD algorithm on a cluster, this will also require a distributed linear solver (see \cite{10.1145/2807591.2807603} and references herein). Efforts are also underway to continue the development of a version where cells are computed in a fully parallel fashion, bypassing the Bowyer-Watson algorithm, and leading naturally to other ways of managing separate diagrams on multiple machines.

\paragraph{Acknowledgements}
The authors wish to thank Roya Mohayaee, Sebastian von Hausegger and Farnik Nikakhtar for testing, using and reporting bugs with the early version of the algorithm. We wish also to thank Rémi Flamary for discussions, as well as the anonymous reviewers for the suggestions they made that helped improving this article.

B. Lévy's work is supported by the COSMOGRAM-launchpad Inria exploratory action (AeX). Q. Mérigot and H. Leclerc acknowledge the support of the Agence nationale de la recherche, through the PEPR PDE-AI project, contract number ANR-23-PEIA-0004.
Experiments presented in this article were carried out using the Grid'5000 testbed, supported by a scientific interest group hosted by Inria and including CNRS, RENATER and several Universities as well as other organizations (see \url{https://www.grid5000.fr})

\bibliographystyle{alpha}
\bibliography{DVD}

\newcommand{\etalchar}[1]{$^{#1}$}
\begin{thebibliography}{dGWH{\etalchar{+}}15}

\bibitem[AHA98]{aurenhammer1998minkowski}
Franz Aurenhammer, Friedrich Hoffmann, and Boris Aronov.
\newblock Minkowski-type theorems and least-squares clustering.
\newblock {\em Algorithmica}, 20:61--76, 1998.

\bibitem[Aur91]{10.1145/116873.116880}
Franz Aurenhammer.
\newblock Voronoi diagrams—a survey of a fundamental geometric data
  structure.
\newblock {\em ACM Comput. Surv.}, 23(3):345–405, sep 1991.

\bibitem[BAR{\etalchar{+}}21]{DBLP:journals/cgf/BasselinARSLL21}
Justine Basselin, Laurent Alonso, Nicolas Ray, Dmitry Sokolov, Sylvain
  Lefebvre, and Bruno L{\'{e}}vy.
\newblock Restricted power diagrams on the {GPU}.
\newblock {\em Comput. Graph. Forum}, 40(2):1--12, 2021.

\bibitem[BB00]{DBLP:journals/nm/BenamouB00}
Jean{-}David Benamou and Yann Brenier.
\newblock A computational fluid mechanics solution to the monge-kantorovich
  mass transfer problem.
\newblock {\em Numerische Mathematik}, 84(3):375--393, 2000.

\bibitem[BC13]{DBLP:journals/dga/BrascoC13}
Lorenzo Brasco and Guillaume Carlier.
\newblock Congested traffic equilibria and degenerate anisotropic pdes.
\newblock {\em Dyn. Games Appl.}, 3(4):508--522, 2013.

\bibitem[BCM24]{BENAMOU2024112745}
J.-D. Benamou, C.J. Cotter, and H.~Malamut.
\newblock Entropic optimal transport solutions of the semigeostrophic
  equations.
\newblock {\em Journal of Computational Physics}, 500:112745, 2024.

\bibitem[BCMO16]{DBLP:journals/nm/BenamouCMO16}
Jean{-}David Benamou, Guillaume Carlier, Quentin M{\'{e}}rigot, and
  {\'{E}}douard Oudet.
\newblock Discretization of functionals involving the monge-amp{\`{e}}re
  operator.
\newblock {\em Numerische Mathematik}, 134(3):611--636, 2016.

\bibitem[BFH{\etalchar{+}}03]{EUR}
Y.~Brenier, U.~Frisch, M.~Hénon, G.~Loeper, S.~Matarrese, R.~Mohayaee, and
  A.~Sobolevskii.
\newblock Reconstruction of the early universe as a convex optimization
  problem.
\newblock {\em Mon. Not. R. Astron. Soc.}, 346(501), 2003.

\bibitem[BFO14]{benamou2014numerical}
Jean-David Benamou, Brittany~D Froese, and Adam~M Oberman.
\newblock Numerical solution of the optimal transportation problem using the
  monge--amp{\`e}re equation.
\newblock {\em Journal of Computational Physics}, 260:107--126, 2014.

\bibitem[Bow81]{DBLP:journals/cj/Bowyer81}
Adrian Bowyer.
\newblock Computing dirichlet tessellations.
\newblock {\em Comput. J.}, 24(2):162--166, 1981.

\bibitem[BPR23]{BOURNE2023104023}
D.P. Bourne, M.~Pearce, and S.M. Roper.
\newblock Geometric modelling of polycrystalline materials: Laguerre
  tessellations and periodic semi-discrete optimal transport.
\newblock {\em Mechanics Research Communications}, 127:104023, 2023.

\bibitem[Bre91]{BrenierPFMR91}
Yann Brenier.
\newblock Polar factorization and monotone rearrangement of vector-valued
  functions.
\newblock {\em Communications on Pure and Applied Mathematics}, 44:375--417,
  1991.

\bibitem[Bre15]{brenier:hal-01137528}
Yann Brenier.
\newblock A double large deviation principle for monge-ampère gravitation.
\newblock working paper or preprint, March 2015.

\bibitem[BY98]{DBLP:books/daglib/0095173}
Jean{-}Daniel Boissonnat and Mariette Yvinec.
\newblock {\em Algorithmic geometry}.
\newblock Cambridge University Press, 1998.

\bibitem[CMYB19]{DBLP:conf/bigdataconf/CaraffaMYB19}
Laurent Caraffa, Pooran Memari, Murat Yirci, and Mathieu Br{\'{e}}dif.
\newblock Tile {\&} merge: Distributed delaunay triangulations for cloud
  computing.
\newblock In Chaitanya~K. Baru, Jun Huan, Latifur Khan, Xiaohua Hu, Ronay Ak,
  Yuanyuan Tian, Roger~S. Barga, Carlo Zaniolo, Kisung Lee, and Yanfang~(Fanny)
  Ye, editors, {\em 2019 {IEEE} International Conference on Big Data {(IEEE}
  BigData), Los Angeles, CA, USA, December 9-12, 2019}, pages 1613--1618.
  {IEEE}, 2019.

\bibitem[CP84]{MC_GEO_1984}
M.~Cullen and R.~Purser.
\newblock An extended lagrangian theory of semi-geostrophic frontogenesis.
\newblock {\em J. of the Atmospheric Sciences}, pages 1477--1497, 1984.

\bibitem[CTP94]{Couchman1994HydraAA}
Hugh M.~P. Couchman, Peter~A Thomas, and Frazer~R. Pearce.
\newblock Hydra: An adaptive--mesh implementation of pppm--sph.
\newblock {\em arXiv: Astrophysics}, 1994.

\bibitem[Cut13]{NIPS2013_af21d0c9}
Marco Cuturi.
\newblock Sinkhorn distances: Lightspeed computation of optimal transport.
\newblock In C.J. Burges, L.~Bottou, M.~Welling, Z.~Ghahramani, and K.Q.
  Weinberger, editors, {\em Advances in Neural Information Processing Systems},
  volume~26. Curran Associates, Inc., 2013.

\bibitem[Dem19]{Demidov2019}
D.~Demidov.
\newblock Amgcl: An efficient, flexible, and extensible algebraic multigrid
  implementation.
\newblock {\em Lobachevskii J. of Math.}, 40(5):535--546, May 2019.

\bibitem[dGWH{\etalchar{+}}15]{DBLP:journals/tog/GoesWHPD15}
Fernando de~Goes, Corentin Wallez, Jin Huang, Dmitry Pavlov, and Mathieu
  Desbrun.
\newblock Power particles: an incompressible fluid solver based on power
  diagrams.
\newblock {\em {ACM} Trans. Graph.}, 34(4):50:1--50:11, 2015.

\bibitem[EBC{\etalchar{+}}22]{EGAN2022111542}
C.P. Egan, D.P. Bourne, C.J. Cotter, M.J.P. Cullen, B.~Pelloni, S.M. Roper, and
  M.~Wilkinson.
\newblock A new implementation of the geometric method for solving the eady
  slice equations.
\newblock {\em Journal of Computational Physics}, 469:111542, 2022.

\bibitem[EGH00]{eymard2000finite}
Robert Eymard, Thierry Gallou{\"e}t, and Rapha{\`e}le Herbin.
\newblock Finite volume methods.
\newblock {\em Handbook of numerical analysis}, 7:713--1018, 2000.

\bibitem[EM90]{10.1145/77635.77639}
Herbert Edelsbrunner and Ernst~Peter M\"{u}cke.
\newblock Simulation of simplicity: A technique to cope with degenerate cases
  in geometric algorithms.
\newblock {\em ACM Trans. Graph.}, 9(1):66–104, jan 1990.

\bibitem[FMMS02]{EURNature}
U.~Frisch, S.~Matarrese, R.~Mohayaee, and A.~Sobolevskii.
\newblock A reconstruction of the initial conditions of the universe by optimal
  mass transportation.
\newblock {\em Nature}, 417(260), 2002.

\bibitem[FSV{\etalchar{+}}19]{feydy2019interpolating}
Jean Feydy, Thibault S{\'e}journ{\'e}, Fran{\c{c}}ois-Xavier Vialard, Shun-ichi
  Amari, Alain Trouve, and Gabriel Peyr{\'e}.
\newblock Interpolating between optimal transport and mmd using sinkhorn
  divergences.
\newblock In {\em The 22nd International Conference on Artificial Intelligence
  and Statistics}, pages 2681--2690, 2019.

\bibitem[GM18]{DBLP:journals/focm/GallouetM18}
Thomas~O. Gallou{\"{e}}t and Quentin M{\'{e}}rigot.
\newblock A lagrangian scheme {\`{a}} la brenier for the incompressible euler
  equations.
\newblock {\em Found. Comput. Math.}, 18(4):835--865, 2018.

\bibitem[Hay20]{hayami2020convergenceconjugategradientmethod}
Ken Hayami.
\newblock Convergence of the conjugate gradient method on singular systems,
  2020.

\bibitem[ILSS06]{DBLP:journals/tog/IsenburgLSS06}
Martin Isenburg, Yuanxin Liu, Jonathan~Richard Shewchuk, and Jack Snoeyink.
\newblock Streaming computation of delaunay triangulations.
\newblock {\em {ACM} Trans. Graph.}, 25(3):1049--1056, 2006.

\bibitem[JKO98]{doi:10.1137/S0036141096303359}
Richard Jordan, David Kinderlehrer, and Felix Otto.
\newblock The variational formulation of the fokker--planck equation.
\newblock {\em SIAM Journal on Mathematical Analysis}, 29(1):1--17, 1998.

\bibitem[KMT19]{KMT2019}
Jun Kitagawa, Quentin Mérigot, and Boris Thibert.
\newblock Convergence of a newton algorithm for semi-discrete optimal
  transport.
\newblock {\em J. Eur. Math. Soc.}, 21(9), 2019.

\bibitem[L\'22]{10.1016/j.jcp.2021.110838}
Bruno L\'{e}vy.
\newblock Partial optimal transport for a constant-volume lagrangian mesh with
  free boundaries.
\newblock {\em J. Comput. Phys.}, 451(C), feb 2022.

\bibitem[LB12]{DBLP:conf/imr/LevyB12}
Bruno L{\'{e}}vy and Nicolas Bonneel.
\newblock Variational anisotropic surface meshing with voronoi parallel linear
  enumeration.
\newblock In Xiangmin Jiao and Jean{-}Christophe Weill, editors, {\em
  Proceedings of the 21st International Meshing Roundtable, {IMR} 2012, October
  7-10, 2012, San Jose, CA, {USA}}, pages 349--366. Springer, 2012.

\bibitem[LBM24]{levy2024monge}
Bruno Lévy, Yann Brenier, and Roya Mohayaee.
\newblock Monge amp\`ere gravity: from the large deviation principle to
  cosmological simulations through optimal transport, 2024.

\bibitem[L{\'{e}}v15]{journals/M2AN/LevyNAL15}
Bruno L{\'{e}}vy.
\newblock A numerical algorithm for {$L_2$} semi-discrete optimal transport in
  3d.
\newblock {\em ESAIM M2AN (Mathematical Modeling and Analysis)}, 2015.

\bibitem[L{\'{e}}v16]{DBLP:journals/cad/Levy16}
Bruno L{\'{e}}vy.
\newblock Robustness and efficiency of geometric programs: The predicate
  construction kit {(PCK)}.
\newblock {\em Comput. Aided Des.}, 72:3--12, 2016.

\bibitem[L{\'{e}}v22]{DBLP:journals/jcphy/Levy22}
Bruno L{\'{e}}vy.
\newblock Partial optimal transport for a constant-volume lagrangian mesh with
  free boundaries.
\newblock {\em J. Comput. Phys.}, 451:110838, 2022.

\bibitem[Lmc23]{WEB:GEOGRAM}
Inria~(Bruno L\'evy and multiple contributors).
\newblock Geogram: a programming library of geometric algorithms.
\newblock \url{http://https://github.com/BrunoLevy/geogram}, 2014-2023.

\bibitem[LMSS20]{DBLP:journals/siamnum/LeclercMSS20}
Hugo Leclerc, Quentin M{\'{e}}rigot, Filippo Santambrogio, and Federico Stra.
\newblock Lagrangian discretization of crowd motion and linear diffusion.
\newblock {\em {SIAM} J. Numer. Anal.}, 58(4):2093--2118, 2020.

\bibitem[LMv21]{10.1093/mnras/stab1676}
Bruno Levy, Roya Mohayaee, and Sebastian von Hausegger.
\newblock {A fast semidiscrete optimal transport algorithm for a unique
  reconstruction of the early Universe}.
\newblock {\em Monthly Notices of the Royal Astronomical Society},
  506(1):1165--1185, 06 2021.

\bibitem[LS18]{DBLP:journals/cg/LevyS18}
Bruno L{\'{e}}vy and Erica~L. Schwindt.
\newblock Notions of optimal transport theory and how to implement them on a
  computer.
\newblock {\em Comput. Graph.}, 72:135--148, 2018.

\bibitem[M{\'e}r11]{DBLP:journals/cgf/Merigot11}
Quentin M{\'e}rigot.
\newblock A multiscale approach to optimal transport.
\newblock {\em Comput. Graph. Forum}, 30(5):1583--1592, 2011.

\bibitem[MFF{\etalchar{+}}17]{mourya2017distributed}
Rahul Mourya, André Ferrari, Rémi Flamary, Pascal Bianchi, and Cédric
  Richard.
\newblock Distributed deblurring of large images of wide field-of-view, 2017.

\bibitem[Mon84]{Monge1784}
Gaspard Monge.
\newblock M\'emoire sur la th\'eorie des d\'eblais et des remblais.
\newblock {\em Histoire de l'Académie Royale des Sciences (1781)}, pages
  666--704, 1784.

\bibitem[MS24]{medina2024flow}
Ismael Medina and Bernhard Schmitzer.
\newblock Flow updates for domain decomposition of entropic optimal transport.
\newblock {\em arXiv preprint arXiv:2405.09400}, 2024.

\bibitem[MT21]{DBLP:journals/corr/abs-2003-00855}
Quentin Merigot and Boris Thibert.
\newblock Optimal transport: discretization and algorithms.
\newblock In {\em Handbook of numerical analysis}, volume~22, pages 133--212.
  Elsevier, 2021.

\bibitem[NPL{\etalchar{+}}23]{PhysRevD.108.083534}
Farnik Nikakhtar, Nikhil Padmanabhan, Bruno L\'evy, Ravi~K. Sheth, and Roya
  Mohayaee.
\newblock Optimal transport reconstruction of biased tracers in redshift space.
\newblock {\em Phys. Rev. D}, 108:083534, Oct 2023.

\bibitem[NSLM22]{nikak_prl_2022}
Farnik {Nikakhtar}, Ravi~K. {Sheth}, Bruno {L{\'e}vy}, and Roya {Mohayaee}.
\newblock {Optimal Transport Reconstruction of Baryon Acoustic Oscillations}.
\newblock {\em Phys. Rev. Lett.}, 129(25):251101, December 2022.

\bibitem[OBSC00]{Okabe00}
Atsuyuki Okabe, Barry Boots, Kokichi Sugihara, and Sung~Nok Chiu.
\newblock {\em Spatial Tessellations: Concepts and Applications of {V}oronoi
  Diagrams}.
\newblock Series in Probability and Statistics. John Wiley and Sons, Inc., 2nd
  ed. edition, 2000.

\bibitem[PC19]{DBLP:journals/ftml/PeyreC19}
Gabriel Peyr{\'{e}} and Marco Cuturi.
\newblock Computational optimal transport.
\newblock {\em Found. Trends Mach. Learn.}, 11(5-6):355--607, 2019.

\bibitem[PMP14]{7013068}
Tom Peterka, Dmitriy Morozov, and Carolyn Phillips.
\newblock High-performance computation of distributed-memory parallel 3d
  voronoi and delaunay tessellation.
\newblock In {\em SC '14: Proceedings of the International Conference for High
  Performance Computing, Networking, Storage and Analysis}, pages 997--1007,
  2014.

\bibitem[PSY{\etalchar{+}}15]{10.1145/2807591.2807603}
Jongsoo Park, Mikhail Smelyanskiy, Ulrike~Meier Yang, Dheevatsa Mudigere, and
  Pradeep Dubey.
\newblock High-performance algebraic multigrid solver optimized for multi-core
  based distributed parallel systems.
\newblock SC '15, New York, NY, USA, 2015. Association for Computing Machinery.

\bibitem[QLdGJ22]{DBLP:journals/tog/QuLGJ22}
Ziyin Qu, Minchen Li, Fernando de~Goes, and Chenfanfu Jiang.
\newblock The power particle-in-cell method.
\newblock {\em {ACM} Trans. Graph.}, 41(4):118:1--118:13, 2022.

\bibitem[RSLL18]{DBLP:journals/tog/RayS0L18}
Nicolas Ray, Dmitry Sokolov, Sylvain Lefebvre, and Bruno L{\'{e}}vy.
\newblock Meshless voronoi on the {GPU}.
\newblock {\em {ACM} Trans. Graph.}, 37(6):265, 2018.

\bibitem[Sam90]{10.5555/77589}
Hanan Samet.
\newblock {\em The design and analysis of spatial data structures}.
\newblock Addison-Wesley Longman Publishing Co., Inc., USA, 1990.

\bibitem[San15]{SantambrogioOTsurvey}
Filippo Santambrogio.
\newblock Optimal transport for applied mathematicians.
\newblock {\em Birk{\"a}user, NY}, 55(58-63):94, 2015.

\bibitem[She97]{shewchuk97a}
Jonathan~Richard Shewchuk.
\newblock Adaptive {P}recision {F}loating-{P}oint {A}rithmetic and {F}ast
  {R}obust {G}eometric {P}redicates.
\newblock {\em Discrete \& Computational Geometry}, 18(3):305--363, October
  1997.

\bibitem[She05]{10.1145/1064092.1064129}
Richard Shewchuk.
\newblock Star splaying: an algorithm for repairing delaunay triangulations and
  convex hulls.
\newblock In {\em Proceedings of the Twenty-First Annual Symposium on
  Computational Geometry}, SCG '05, page 237–246, New York, NY, USA, 2005.
  Association for Computing Machinery.

\bibitem[Slo87]{SLOAN198734}
S.W. Sloan.
\newblock A fast algorithm for constructing delaunay triangulations in the
  plane.
\newblock {\em Advances in Engineering Software (1978)}, 9(1):34--55, 1987.

\bibitem[Spr10]{10.1111/j.1365-2966.2009.15715.xISTEX}
Volker Springel.
\newblock {E pur si muove: Galilean-invariant cosmological hydrodynamical
  simulations on a moving mesh}.
\newblock {\em Monthly Notices of the Royal Astronomical Society},
  401(2):791--851, 01 2010.

\bibitem[{The}23]{cgal:eb-23b}
{The CGAL Project (multiple authors)}.
\newblock {\em {CGAL} User and Reference Manual}.
\newblock {CGAL Editorial Board}, {5.6} edition, 2023.

\bibitem[Vil03]{opac-b1122739}
Cédric Villani.
\newblock {\em Topics in optimal transportation}.
\newblock Graduate studies in mathematics. American Mathematical Society,
  Providence (R.I.), 2003.

\bibitem[Vil09]{OTON}
Cédric Villani.
\newblock {\em Optimal transport : old and new}.
\newblock Grundlehren der mathematischen Wissenschaften. Springer, Berlin,
  2009.

\bibitem[vLM22]{vhauss_prl_2022}
Sebastian {von Hausegger}, Bruno {L{\'e}vy}, and Roya {Mohayaee}.
\newblock {Accurate Baryon Acoustic Oscillations Reconstruction via
  Semidiscrete Optimal Transport}.
\newblock {\em Phys. Rev. Lett.}, 128(20):201302, May 2022.

\bibitem[Wat81]{journals/cj/Watson81}
David Watson.
\newblock Computing the n-dimensional delaunay tessellation with application to
  voronoi polytopes.
\newblock {\em Comput. J.}, 24(2):167--172, 1981.

\end{thebibliography}
\end{document}